%% file: main.tex
\documentclass{article}

\usepackage[utf8]{inputenc}
\usepackage{tismir}
\usepackage{amsmath}
\usepackage{amssymb}
\usepackage[hidelinks]{hyperref}
\usepackage{url}
\usepackage{graphicx}
\usepackage{ifthen}
\usepackage{booktabs}
\usepackage{lipsum}
\usepackage{acronym}
\usepackage{multirow}
\usepackage{pifont}
\usepackage{siunitx}
\usepackage{subcaption}
\usepackage[table]{xcolor}
\usepackage{float}

\hypersetup{
    colorlinks=true,
    linkcolor=red,
    citecolor=magenta,
    filecolor=magenta,      
    urlcolor=blue
    }

\acrodef{EMA}{Exponential Moving Average}
\acrodef{SCE}{Shifted Cross-Entropy}
\acrodef{VQT}{Variable-$Q$ Transform}

\newboolean{usetodo}
\newboolean{rebuttal}
\setboolean{usetodo}{false}
\setboolean{rebuttal}{false}

\ifthenelse{\boolean{usetodo}}
{
    \newcommand{\TODO}[1]{{\color{red}TODO: #1}}
    \newcommand{\alain}[1]{{\color{green}al1: #1}}
    \newcommand{\bernardo}[1]{{\color{blue}bern: #1}}
    \newcommand{\stefan}[1]{{\color{blue}sl: #1}}
    \newcommand{\ben}[1]{{\color{brown}ben: #1}}
}{
    \newcommand{\TODO}[1]{}
    \newcommand{\alain}[1]{}
    \newcommand{\bernardo}[1]{}
    \newcommand{\stefan}[1]{}
    \newcommand{\ben}[1]{}
}
\ifthenelse{\boolean{rebuttal}}
{
    \newcommand{\new}[1]{{\color{blue}#1}}
    \newcommand{\morenew}[1]{{\color{magenta}#1}}
}{
    \newcommand{\new}[1]{#1}
    \newcommand{\morenew}[1]{#1}
}

\newcommand{\RR}{\mathbb{R}}
\newcommand{\ZZ}{\mathbb{Z}}

\newcommand{\GG}{\mathcal{G}}

\newcommand{\XX}{\mathcal{X}}
\newcommand{\YY}{\mathcal{Y}}

\newcommand{\cmark}{\ding{51}}%
\newcommand{\xmark}{\ding{55}}%

\renewcommand{\b}[1]{\mathbf{#1}}
\newcommand{\bk}[1]{\mathbf{#1}^{(k)}}
\newcommand{\bs}[1]{\widetilde{\mathbf{#1}}}
\newcommand{\sk}[1]{\widetilde{#1}^{(k)}}
\newcommand{\bsk}[1]{\widetilde{\mathbf{#1}}^{(k)}}
\newcommand{\SNR}{\text{SNR}}

\newcommand{\kmax}{k_{\max}}
\newcommand{\x}{\b{x}}
\newcommand{\y}{\b{y}}

\DeclareMathOperator*{\argmax}{arg\,max}
\DeclareMathOperator*{\sg}{sg}

\newcommand{\LL}[1]{\mathcal{L}_{\text{#1}}}

\acrodef{RPA}{Raw Pitch Accuracy}
\acrodef{RCA}{Raw Chroma Accuracy}
\acrodef{MnE}{Mean Pitch Error}
\acrodef{MdE}{Median Pitch Error}

\acrodef{CQT}{Constant-$Q$ Transform}
\acrodef{DDSP}{Differentiable Digital Signal Processing}
\acrodef{MIR}{Music Information Retrieval}
\acrodef{SSL}{Self-Supervised Learning}

\title{PESTO: Real-Time Pitch Estimation with Self-supervised Transposition-equivariant Objective}
\author{%
    Alain Riou\thanks{Equal contribution}, %
    Bernardo Torres\protect\footnotemark[1], %
    Ben Hayes, %
    Stefan Lattner,\\%
    Gaëtan Hadjeres, %
    Gaël Richard, %
    Geoffroy Peeters\\%
    {\small\protect\footnotemark[1] Equal contribution}
    \vspace{-3mm}
}
\date{}

\type{research}

\authorref{Riou,~A., Torres,~B., Hayes,~B., Lattner,~S., Hadjeres,~G., Richard,~G., and Peeters,~G.}
\journalyear{2025}
\journalvolume{8}
\journalissue{1}
\journalpages{334--352}
\doi{10.5334/TISMIR.251}

\authorshort{Riou, Torres, et al.} %or, e.g., \authorshort{Author1 et al}
\begin{document}

%%%%%%%%%%%%%%%%%%%%%%%%%%%%%%%%%%%%%%%%%%%%%%%%%%%%%%%%%%%%%%%%%%%%%%%%%%%%%%%%
% Abstract
%%%%%%%%%%%%%%%%%%%%%%%%%%%%%%%%%%%%%%%%%%%%%%%%%%%%%%%%%%%%%%%%%%%%%%%%%%%%%%%%

\twocolumn[{%
\maketitleblock
\begin{abstract}
    In this paper, we introduce PESTO, a self-supervised learning approach for single-pitch estimation using a Siamese architecture. Our model processes individual frames of a Variable-$Q$ Transform (VQT) and predicts pitch distributions. The neural network is designed to be equivariant to translations, notably \new{thanks to} a Toeplitz fully-connected layer. In addition, we construct pitch-shifted pairs by translating and cropping the VQT frames and train our model with a novel class-based transposition-equivariant objective, eliminating the need for annotated data. Thanks to this architecture and training objective, our model achieves remarkable performances while being very lightweight ($130$k parameters).

    Evaluations on music and speech datasets (MIR-1K, MDB-stem-synth, and PTDB) demonstrate that PESTO not only outperforms self-supervised baselines but also competes with supervised methods, exhibiting superior cross-dataset generalization. Finally, we enhance PESTO's practical utility by developing a streamable VQT implementation using cached convolutions. Combined with our model's low latency (less than \SI{10}{\milli\second}) and minimal parameter count, this makes PESTO particularly suitable for real-time applications.
\end{abstract}
}
]
%\saythanks{blablalbal}

%%%%%%%%%%%%%%%%%%%%%%%%%%%%%%%%%%%%%%%%%%%%%%%%%%%%%%%%%%%%%%%%%%%%%%%%%%%%%%%%
% Main Content Start
%%%%%%%%%%%%%%%%%%%%%%%%%%%%%%%%%%%%%%%%%%%%%%%%%%%%%%%%%%%%%%%%%%%%%%%%%%%%%%%%

\section{Introduction}

Pitch, alongside loudness and timbre, is one of the primary auditory sensations and plays a central role in audio perception. Its automatic estimation is a fundamental task in audio analysis, with broad applications in music information retrieval (MIR) and speech processing.

Pitch perception is closely tied to the acoustic periodicity of the audio waveform. \new{From a spectral perspective, it is related to the spacing between its partials that corresponds to the fundamental frequency (F0)~\citep{Yost2009,Oxenham2012}.} In the case of harmonic or quasi-harmonic signals, these two characteristics coincide, so that the perceived pitch is often measured via F0 estimation.
This has been a long-standing research problem, with numerous solutions based on signal processing techniques being proposed~\citep{Noll1967,ACF,YIN,SWIPE}.

More recently, deep learning methods such as CREPE~\citep{CREPE} have achieved almost perfect results, so pitch estimation is often considered a solved problem. However, these approaches exhibit several limitations: they are not lightweight, rely heavily on annotated data that is difficult to obtain, and fail to generalize well to out-of-domain data, limiting their practical utility. Additionally, they rely on non-causal convolutional neural networks with large kernels applied to downsampled audio waveforms, making them unsuitable for real-time applications.
%Although self-supervised methods \citep{SPICE,DDSPinv} have been proposed to address pitch estimation without annotated data, their performance remains significantly below that of supervised baselines.

Real-time estimation of fundamental frequency is, indeed, an integral component of a range of speech and music audio systems. These include staples of modern audio production, such as pitch correction and audio-to-MIDI conversion, as well as creative musical effects such as vocal harmonisers. Interactive digital tools for music education\footnote{e.g. \href{https://yousician.com/}{Yousician}, \href{https://developer.sony.com/spresense/sme-use-cases}{Yuru}}, creativity support tools, and jam-along systems also rely heavily on real-time inference of musical attributes including pitch~\citep{Meier2023}.
As noted by \citet{stefani_challenges_2022}, real-time application of MIR algorithms is generally challenging due to computational limitations and the availability of only causal information.
Selecting appropriate algorithms for pitch tracking thus typically requires a compromise between efficiency and performance.

Neural audio models have enabled new categories of audio processing, such as timbre transfer \citep{huang_timbretron_2018-1}, speaking voice conversion \citep{sisman_overview_2021-1, bargum_reimagining_2024}, and singing voice conversion \citep{huang_singing_2023}. These techniques have garnered interest in real-time applications, as demonstrated by various open- and closed-source software releases\footnote{See, for example, \href{https://mawf.io/}{Mawf}, \href{https://magenta.tensorflow.org/ddsp-vst}{DDSP-VST}, and \href{https://neutone.ai/}{Neutone}.}.

Differentiable digital signal processing (DDSP) \citep{ENGEL-DDSP}, a hybrid of deep learning and signal processing, has been central to these advances \citep{hayes_review_2023}. %DDSP typically uses a neural network to parameterize a synthesizer or signal processor.
Due to the challenges of optimizing oscillation frequency by gradient descent \citep{hayes_sinusoidal_2023}, such models often rely on fundamental frequency (F0) estimates to predict synthesis parameters and drive oscillatory components.

Neural source-filter models \citep{wang_neural_2019-2}, another class of hybrid neural vocoders, employ periodic excitation signals with F0 matching the target speech, and are widely used in singing voice conversion (SVC) \citep{huang_singing_2023}. Popular real-time SVC systems, such as So-VITS-SVC\footnote{\url{https://github.com/svc-develop-team/so-vits-svc}} and DDSP-SVC\footnote{\url{https://github.com/yxlllc/DDSP-SVC}}, depend on external F0 estimators, which contribute to inference latency.

For real-time applications of such models, there is a tension between what they require from an F0 estimator at training time, i.e., high-quality F0 annotations, and at inference time, i.e., streaming prediction with low latency. As a result, practitioners face a challenging design choice. Some opt to use an efficient estimator for both training and inference. For example, some applications \citep{yang_streamvc_2024, ganis_real-time_2021, fabbro_speech_2020} opt for a classical method such as Yin \citep{YIN}, while others use a knowledge-distilled neural network\footnote{As, for example, in the \href{https://magenta.tensorflow.org/ddsp-vst}{DDSP-VST} plugin.}. This minimises inference latency, at the expense of training with sub-optimal F0 annotations which may harm performance \citep{Alonso2021}. Others opt to use a high-quality neural pitch estimator, which will provide more reliable annotations at the expense of considerable inference latency. Alternatively, practitioners may decide to train with the high-quality estimator and substitute in the more efficient algorithm at inference time, at the risk of introducing a mismatch between train and test conditions.

\paragraph{Paper proposal.}
This work addresses such training and inference design trade-offs by proposing several key contributions: (i) a lightweight state-of-the-art neural pitch estimator, (ii) a self-supervised training strategy which does not require labeled data, (iii) a real-time causal inference model.
% Such design trade-offs would be alleviated by a pitch estimator which: (i) achieves real-time inference, (ii) introduces zero structural latency (i.e. is entirely causal), and (iii) achieves comparable performance to state-of-the-art neural F0 estimators. Our proposed model achieves all these aims, and does so with a self-supervised training method which obviates the need for annotated data.% \TODO{nice transition}

%In this work, we propose a new self-supervised method for pitch estimation that requires no annotated data.
More precisely,  our approach applies a \ac{VQT} to audio signals and processes pitch-shifted frames using a Siamese architecture (Section~\ref{sec:frontend}). We incorporate a Toeplitz fully-connected layer into the architecture, enabling it to naturally preserve transpositions (Section~\ref{sec:architecture}). Furthermore, we formulate pitch estimation as a classification problem and introduce a novel loss function that enforces pitch-shift equivariance without requiring a decoder (Section~\ref{sec:objective}).
To further enhance practical utility, we propose a simple yet effective modification that allows the model to process audio streams in real-time (Section~\ref{sec:real_time}).

We evaluate our model both on music and speech datasets (Section~\ref{sec:experimental_setup}). Our results (Section~\ref{sec:results}) indicate that it significantly outperforms prior self-supervised baselines and achieves results competitive with supervised methods, using $68$ times fewer parameters than the most lightweight supervised model. In particular, we observe better generalization to unseen datasets compared to the baselines.

Moreover, because our model operates on VQT frames rather than raw audio waveforms, it can handle arbitrary sampling rates and hop sizes during inference without requiring downsampling, regardless of the parameters used during training.
These features, combined with the ability to retrain or fine-tune the model without annotated data and minimal computational resources, make it highly practical for various real-world scenarios.

To facilitate its usage and encourage further research in this direction, we open-source the full training code and release a pip-installable package along with pretrained models.\footnote{\url{https://github.com/SonyCSLParis/pesto}}

This paper extends our previous publication~\citep{PESTO} in several ways.
We replace the original frontend (CQT) by a VQT and show it drastically improves performances. We also introduce slight architectural changes.
Moreover, we provide a deeper analysis with additional experiments, extending evaluations to speech data and comparing with the recent PENN model~\citep{PENN}. More detailed ablation studies are also conducted.
In addition, we implement and release a real-time implementation of our model compatible with audio streams, enhancing its practical utility. Finally, we improve the clarity of the paper with refined explanations, rebranded figures, and more detailed discussions.

\section{Related work}

\subsection{Pitch estimation}

Monophonic pitch estimation has been a subject of interest for over fifty years~\citep{Noll1967}. 
The earlier methods typically obtain a pitch curve by processing a candidate-generating function such as cepstrum~\citep{Noll1967}, autocorrelation function~\citep{ACF}, and average magnitude difference function~\citep{AMDF}.
Other functions, such as the normalized cross-correlation function (NCCF)~\citep{PRAAT,RAPT} and the cumulative mean normalized difference function~\citep{YIN,pYIN}, have also been proposed.
On the other hand, \citet{SWIPE} perform pitch estimation by predicting the pitch of the sawtooth waveform whose spectrum best matches the one of the input signal.

As in many other domains, these signal processing-based approaches have been recently outperformed by data-driven ones. In particular, CREPE~\citep{CREPE} is a deep neural network consisting of six convolutional blocks followed by a fully connected layer. Operating directly on waveform data, CREPE processes audio chunks of 64 ms (1024 samples at 16 kHz). Trained in a supervised manner on a large collection of music datasets, it achieves exceptional performance and has become a widely adopted solution for monophonic pitch estimation.

Building on CREPE, \citet{DeepF0} introduce dilated convolutions with residual connections to expand the receptive field, while \citet{FCNF0} improve the architecture further and propose downsampling input signals to \SI{8}{\kilo\hertz} to reduce computational costs. \citet{PENN} later proposed additional training strategies, such as incorporating layer normalization, increasing batch sizes, and making minor architectural changes to enhance performance.

These advancements have led to state-of-the-art results on both music and speech datasets. However, the fundamental framework—modeling pitch estimation as a supervised classification problem—remains unchanged. This dependence on large quantities of annotated data is a significant drawback, as obtaining precise F0 annotations is time-consuming and challenging. Furthermore, these methods often perform poorly when applied to data outside the training distribution~\citep{PENN}, limiting their applicability in diverse scenarios.

\subsection{Self-supervised learning with Siamese networks}

\ac{SSL} has emerged as a promising paradigm to address the challenges associated with gathering large quantities of annotated data, which is often tedious, error-prone, and biased. SSL leverages the data itself to provide a supervision signal by training a neural network to solve a \emph{pretext task}. One common pretext task involves training Siamese networks~\citep{Siamese} to project data points into a latent space and minimize the distance between pairs of inputs that share semantic information. These \emph{positive samples} (or views) can be artificially created by applying transforms to an input data point.

In other words, the goal of Siamese networks is to learn a mapping $f : \mathcal{X} \to \mathcal{Y}$ that is \emph{invariant} to a set of transforms $\mathcal{T} : \mathcal{X} \to \mathcal{X}$. That is, for any data point $\x \in \mathcal{X}$ and any transform $t \in \mathcal{T}$, the following holds:
\begin{equation} f(t(\x)) = f(\x). \label{eq:invariance} \end{equation}

The set of transforms $\mathcal{T}$ is usually composed of semantic-preserving data augmentations. For example, in the image domain, \citet{SimCLR} propose using transforms such as cropping, rotation, and color jittering, which drastically change the pixel values but preserve the content of the image itself. Later, \citet{BAEVSKI-DATA2VEC} propose creating views by masking the input.

However, without additional constraints, the network might simply learn the trivial solution to Eq.\eqref{eq:invariance}, mapping all inputs to the same point.
This phenomenon, called \emph{representation collapse}, can be typically prevented by adding new loss terms computed over several data points. Successful approaches,  typically requiring large batch sizes, include minimizing the similarity between negative samples through a contrastive loss \citep{SimCLR}, or regularizing over the batch statistics~\citep{Wang2020a,BarlowTwins,VICReg}.

% This phenomenon, called \emph{representation collapse}, can be prevented by maximizing the similarity within pairs of positive samples while minimizing the similarity between negative samples through a contrastive loss\citep{SimCLR}. 

% Other approaches involve directly optimizing a regularization loss over the batch statistics~\citep{Wang2020a,BarlowTwins,VICReg}.
% In practice, they have been shown to be equivalent to contrastive ones~\citep{Garrido2023}, and both typically require large batch sizes.

Introducing asymmetry in the Siamese architecture can also be exploited to prevent collapse by changing the training dynamics, such as adding a predictor network after one branch while stopping gradients in the other~\citep{BYOL,SimSiam}. This approach relies solely on positive pairs and is less sensitive to batch size.

\subsection{Equivariant self-supervised learning}

Most \ac{SSL} models aim to learn a mapping that is \emph{invariant} to certain transformations. However, recent research has explored the idea of learning \emph{equivariant} mappings instead.% In fact, full invariance to data augmentations may sometimes harm performance in downstream tasks. \stefan{that's not a good argument for equivariant training, as the applications and use cases are very different.}\alain{that's true, I remove this sentence}\footnote{For instance, invariance to color jittering can negatively impact classification tasks such as flower or bird recognition, as shown in prior work (e.g., \citet{Lee2021}).}

Equivariance, in contrast to invariance, can be mathematically expressed as follows. Let $\GG$ be a group, with group actions $t : \XX \times \GG \to \XX$ and $t' : \YY \times \GG \to \YY$. A mapping $f : \XX \to \YY$ is said to be equivariant with respect to $\GG$ if, for any $x \in \XX$ and $g \in \GG$, the following condition holds:
\begin{equation}
    f(t(\x, g)) = t'(f(\x), g),
    \label{eq:equivariance}
\end{equation}
In other words, $t$ and $t'$ are transforms acting on the input and output spaces, respectively, with parameters from the group $\GG$, and a mapping $f$ is equivariant with regard to $\GG$ if transforming the input $\x$ by $t$ results in a corresponding transformation of the output $f(\x)$ under $t'$.

Note that invariance is a special case of equivariance, where $t'(\cdot, g)$ acts as the identity function for all $g \in \GG$. However, in general, Eq.~\eqref{eq:equivariance} does not have a trivial solution for $f$. As a result, Siamese networks trained to optimize an equivariance objective are not prone to collapse.

Equivariant representation learning has been explored as a means to navigate the latent space of (variational) autoencoders~\citep{Hinton2011, Falorsi2018}. More recently, efforts have been made to incorporate equivariance into self-supervised models, conditioning parts of the neural architecture on the parameters of the transforms applied~\citep{Dangovski2021, EquiMod}. These approaches have shown particular promise in computer vision tasks that involve rotations~\citep{Winter2022, SIE}.

\subsection{Self-supervised learning for MIR}

While originally developed for image representation learning, the aforementioned  approaches have been successfully adapted to the audio domain~\citep{COLA,AudioBarlowTwins,BYOLA}. Notably, contrastive pretraining has demonstrated impressive results for various music information retrieval (MIR) downstream tasks, such as auto-tagging and genre classification~\citep{CLMR,MULE}.% \TODO{some others?}

Few studies have attempted to train neural networks which have equivariant properties to pitch and tempo shifts.
SPICE~\citep{SPICE} offers a method for pitch estimation without the need for annotated data. It creates pairs of views by pitch-shifting an input \ac{CQT} frame by a known number of semitones, $k_1$ and $k_2$. The model learns a mapping $f$ that projects a CQT frame to a scalar value, and it is trained to ensure that the difference between the scalar projections is proportional to the pitch difference, $k_2 - k_1$.
However, \citet{SPICE} observe that this objective alone is insufficient for training, and thus the model requires an additional decoder and a reconstruction loss to prevent collapse. This approach has been extended to tempo~\citep{Morais2023, Gagnere2024} and key estimation \citep{kongSTONESelfsupervisedTonality2024}.

\citet{Quinton2022} applies a comparable approach to tempo estimation. In this case, an audio segment is time-stretched by two factors, $\alpha_1$ and $\alpha_2$, to create a pair of segments $(x_1, x_2)$. A neural network then projects these segments to scalars $z_1$ and $z_2$, and the model is trained to minimize the following loss function:
\begin{equation}
    \LL = \left| \dfrac{z_2}{z_1} - \dfrac{\alpha_2}{\alpha_1} \right|.
\end{equation}

This objective encourages the scalar projections to be proportional to the actual tempo of the song. Unlike SPICE, \citet{Quinton2022} does not require a decoder, and the model avoids collapse by using only the scalar projection loss.

Another family of self-supervised and unsupervised methods for pitch estimation rely on analysis-by-synthesis. \cite{DDSPinv} train a neural network to predict the parameters of a differentiable harmonic-plus-noise synthesizer that approximates the input audio, relying on parameter regression on synthetically generated signals. On a similar approach,  \cite{TORRES-ICASSP} suggests instead using an optimal transport inspired spectral loss function for training a fundamental frequency estimator for harmonic signals.

\section{PESTO}

Our model is a neural network $f_{\theta} : \RR^{F'} \to [0, 1]^K$, with parameters $\theta$, which takes as input a truncated \ac{VQT} frame and returns a pitch distribution $\y = (y_1, \dots, y_K) \in [0, 1]^K$.

\subsection{Frontend}\label{sec:frontend}

In this section, we first describe the Constant-Q Transform (CQT), which serves as a specification for the Variable-Q Transform (VQT) used in our experiments. We then introduce the VQT, which is a variant of the CQT which smoothly decreases the Q factors of the analysis filters for low frequencies. 

\subsubsection{The Constant-Q Transform (CQT)}

The Constant-Q Transform (CQT) shares fundamental principles with wavelet analysis, offering a frequency-dependent time-frequency resolution~\citep{brownCalculationConstantSpectral1991}. Its defining characteristic is the constant ratio (Q) between center frequency and bandwidth across all frequency bins. The center frequencies $f_k$ follow a geometric progression:
\begin{equation}
f_k = f_\text{min} 2^{\frac{k}{12 * B}},
\end{equation}

\noindent where $f_\text{min}$ represents the lowest analysis frequency and $B$ denotes the number of bins per semitone in an equal-tempered scale. This logarithmic frequency spacing creates a musically intuitive representation with the property that in its log-frequency domain, a frequency translation represents a chromatic transposition regardless of the absolute frequency while maintaining the distance between harmonics for a given pitch.
 In practical implementations, particularly in MIR, the CQT typically employs a uniform hop size across all frequency bands matched to the highest frequency's requirements. While this approach introduces redundancy in the low frequency bands compared to the Wavelet transform, it ensures temporal alignment of CQT coefficients across all frequency bins, facilitating subsequent analysis and processing tasks.
%\TODO{triple check if this explanation is correct and clear enough}.

\begin{figure}
    \centering
    \includegraphics[width=\linewidth]{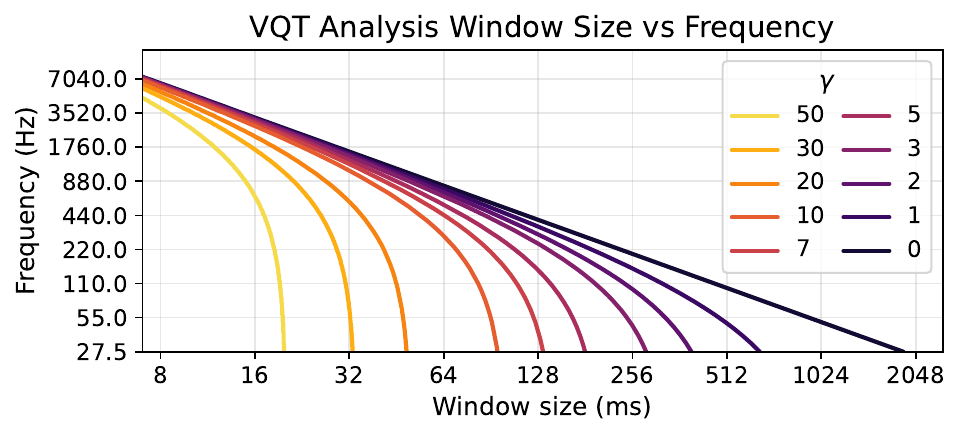}
    \caption{VQT analysis window size as a function of $\gamma$ and analysis center frequency. During the computation of the VQT, all kernels are padded to the nearest power of $2$ to the largest analysis window. The CQT corresponds to the case $\gamma=0$. Both axes are log-scaled.}
    \label{fig:vqt}
\end{figure}

\subsubsection{The Variable-Q Transform (VQT)}

The Variable-$Q$ Transform (VQT) \citep{schorkhuberMatlabToolboxEfficient2014} \morenew{differs from the CQT by adding} a parameter $\gamma$, which smoothly decreases the Q factors of the analysis filters for low frequencies. For \morenew{center} frequency $f_k$, the length of the analysis window is given by
\begin{equation}\label{eq:window}
    w_k = \left\lceil \dfrac{Q f_s}{f_k + \frac{\gamma}{\zeta}} \right\rceil,
\end{equation}
where $Q$ is the Q factor of the analysis filters, $f_s$ is the sampling rate, and \new{$\zeta = 2^{\frac{1}{12 B}} - 1$} is a constant that depends on \morenew{$B$, the number of bins per semitone}, following the implementations of \citet{schorkhuberMatlabToolboxEfficient2014, nnAudio}. For $\gamma=0$, the VQT is equivalent to the CQT. As in the CQT, the \morenew{center} frequencies $\{f_k\}_{k=1}^K$ follow an exponential scaling. Figure \ref{fig:vqt} illustrates the analysis window size as a function of $\gamma$ and \morenew{center} frequency.

% alpha = 2.0 ** (1.0 / bins_per_octave) - 1.0
%     lengths = np.ceil(Q * fs / (freqs + gamma / alpha))

Since the time-domain analysis window for low frequencies is smaller than that of the CQT, the VQT can be computed more efficiently in terms of both time and memory. As we demonstrate empirically, smaller analysis windows that do not span several frames are also beneficial for the pitch estimation task. 

\subsubsection{Pitch-shift in the VQT domain}\label{sec:vqt_shift}

\begin{figure}
    %\centering
    \includegraphics[width=0.45\textwidth]{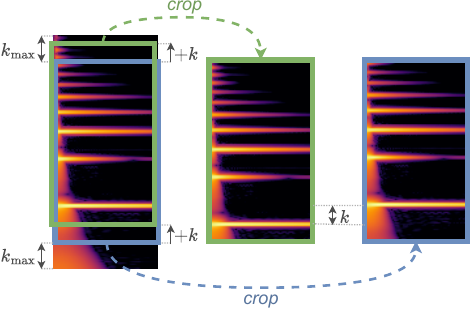}
    \caption{Illustration of the pitch-shift process in the VQT domain. From a given frame, we construct two views by cropping equally-sized sub-frames from it, with a shift of $k$ between them. Since the frequency scale is logarithmic in the VQT domain, this translation corresponds to a\morenew{n approximate}  pitch shift of $k$ bins.}
    \label{fig:cqt}
\end{figure}

Our technique to build pitch-shifted views from a single VQT frame, originally proposed for the CQT by \citet{SPICE}, is depicted in Figure~\ref{fig:cqt}.

More precisely, from a full VQT frame $(x_1, \dots, x_F)$, an integer $k_{\max} \leq F/2$, let $F' = F - 2 k_{\max}$.
Then, given an integer $k$ sampled from $\left[ -\kmax, \kmax \right]$, one can easily compute frames with approximately the same timbre and a pitch shift of $k$ bins\morenew{, i.e. $k/B$ semitones,} by extracting
$$\x = \left(x_{\kmax}, \dots, x_{F - \kmax}\right) \in \RR^{F'}$$
and
$$\bk{x} = \left(x_{\kmax + k}, \dots, x_{F - \kmax + k}\right) \in \RR^{F'}$$
from the original VQT frame $\x$. We set $\kmax$ to $16$. 

\begin{figure*}
    \centering
    \includegraphics[width=\textwidth]{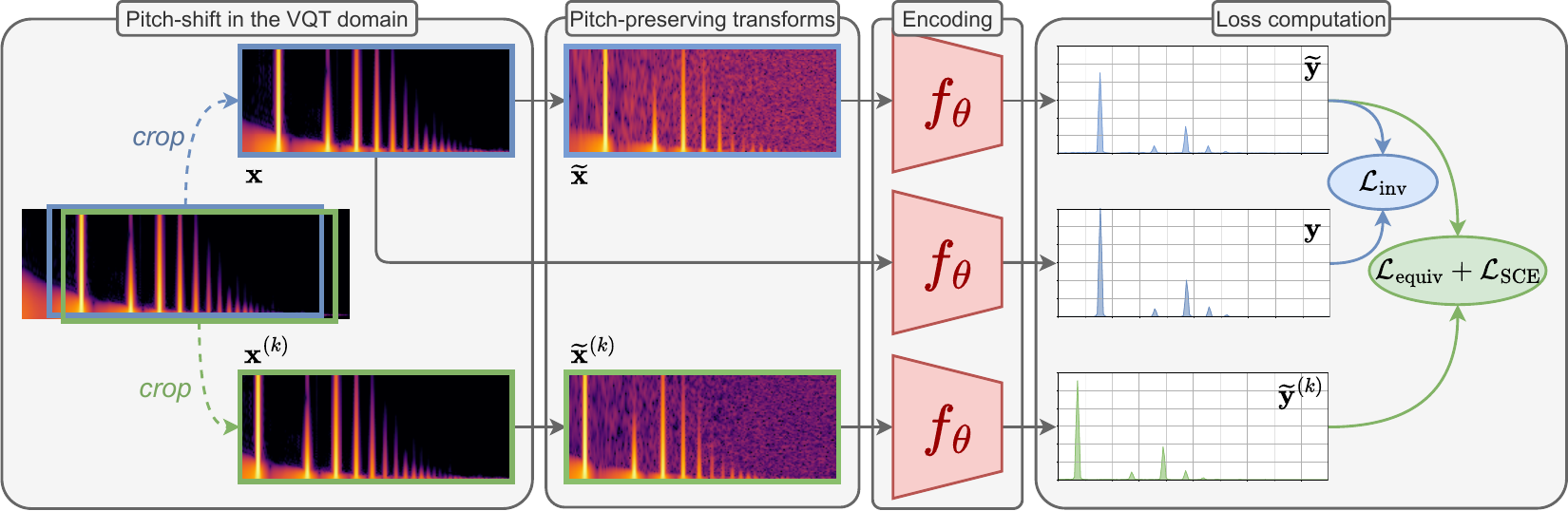}
    \caption{Overview of the PESTO model. Given a \new{1D} VQT frame \morenew{(displayed horizontally, where the horizontal axis corresponds to frequency)}, we first crop it as described in section \ref{sec:vqt_shift} to create a pair of pitch-shifted views $(\x, \bk{x})$. We then obtain $\bs{x}$ and $\bsk{x}$ by randomly applying pitch-preserving transforms to the views. The neural network $f_{\theta}$ predicts pitch distributions from the different views and is trained by minimizing both an invariance loss between $\y$ and $\bs{y}$ and an equivariance loss between $\bs{y}$ and $\bsk{y}$. }
    \label{fig:model}
\end{figure*}    

\subsection{Training objective}\label{sec:objective}

The training procedure for our model takes inspiration from previous SSL methods based on Siamese networks. It is depicted in Figure \ref{fig:model}.

Given a single VQT frame $\x$, we first crop the frame with a known offset and then \morenew{translate} it by a random but known number of bins $k$. Let $\bk{x}$ be the resulting pitch-shifted frame.
Then, we randomly apply a set of \emph{pitch-preserving} transformations to both $\x$ and $\bk{x}$ \new{(white noise, random gain)}. We denote the transformed versions as $\bs{x}$ and $\bsk{x}$, respectively.
Next, we pass $\x$, $\bs{x}$, and $\bsk{x}$ through the same neural network \new{$f_{\theta} : \RR^{F'} \to [0, 1]^K$}, obtaining the pitch distributions $\b{y}$, $\bs{y}$, and $\bsk{y}$, respectively.

We know by construction that $\x$ and $\bs{x}$ have the same pitch. Therefore, we want their pitch distributions $\b{y}$ and $\bs{y}$ to be equal, ensuring that $f_{\theta}$ is \emph{invariant} to the transformations.
Similarly, we know that the pitch difference between $\bs{x}$ and $\bsk{x}$ is exactly $k$ bins. Thus, \morenew{we want their pitch distributions to be} equal but shifted by $k$ bins, ensuring that the network is \emph{equivariant} to the pitch shift.

We train the network $f_{\theta}$ to minimize objectives that enforce these invariance and equivariance constraints.

\subsubsection{Invariance loss}

We want the distributions $\b{y}$ and $\bs{y}$ to be as close as possible. To achieve this, we simply minimize the cross-entropy of $\b{y}$ relative to $\bs{y}$:
\begin{equation}
    \LL{inv}(\b{y}, \bs{y}) = \sum_{i=1}^K \widetilde{y}_i \log y_i.
\end{equation}

\subsubsection{Equivariance loss}

By construction, there is a pitch shift of exactly $k$ bins between $\bs{x}$ and $\bsk{x}$. In other words, for \emph{any} pitch $k' \in \{1, \dots, K\}$, the probability for the pitch of $\bs{x}$ to be $k'$ equals the probability for the pitch of $\bsk{x}$ to be $k' + k$. For example, if we apply a pitch shift of +1 semitone to the frame, the probability that the original frame is a C4 is equal to the probability that the transposed one is a C\#4, and it holds for every pitch independently of the ground truth pitch of the frame.

As $f_\theta$ should return the pitch distribution of its input, we therefore want its outputs $\bs{y}$ and $\bsk{y}$ to verify\stefan{one sentence intuitive explanation why we do this and why it works would be nice}:
\begin{equation}
    \label{eq:cond}
    \widetilde{y}_{k' + k}^{(k)} = \widetilde{y}_{k'},
\end{equation}
for all $k' \in \{1, \dots, K - k\}$.

To achieve this, we consider
$$\b{a} = \left( \alpha, \alpha^2, \dots, \alpha^K \right) \in \RR^K,$$
where $\alpha > 0$ is a fixed scalar.

Then, assume that
\begin{equation}
    \label{eq:condzero}
    \begin{cases}
        \sk{y}_{k'} = 0 &\text{ for all } k' \leq k \\
        \widetilde{y}_{k'} = 0 &\text{ for all } k' > K-k
    \end{cases}
\end{equation}
If Eqn.~\eqref{eq:cond} holds, then
\begin{equation}
    \label{eq:res}
    \begin{aligned}
        \b{a} \cdot \bsk{y}
        &= \sum_{i=k+1}^K \alpha^i \sk{y}_i \\
        %&= \sum_{i=1}^{K-k} \alpha^{i+k} \sk{y}_{i+k} \\
        &= \sum_{i=1}^{K-k} \alpha^{i+k} \widetilde{y}_i & \text{by Eqn. \eqref{eq:cond}}  \\
        %&= \alpha^k \sum_{i=1}^{K-k} \alpha^i \widetilde{y}_i \\
        &= \alpha^k \b{a} \cdot \bs{y}
    \end{aligned}
\end{equation}

Taking inspiration from \citet{Quinton2022}, we therefore define our criterion $\LL{equiv}$ as
\begin{equation}
    \LL{equiv} \left( \bs{y}, \bsk{y}, k \right) = h_\tau \left( \dfrac{\b{a} \cdot \bsk{y}}{\b{a} \cdot \bs{y}} - \alpha^k \right),
\end{equation}
where $h_\tau$ stands for the Huber loss function~\citep{Huber1964}, defined for any scalar $x \in \RR$ by
\begin{equation}
    h_\tau(x) =
    \begin{cases}
        \frac{x^2}{2} & \text{ if } |x| \leq \tau \\
        \frac{\tau^2}{2} + \tau(|x| - \tau) & \text{ otherwise}
    \end{cases}.
\end{equation}
\new{Thanks to $h_\tau$, the gradients of our equivariance loss are proportional to the error when it is small enough (e.g., a few cents) but constant otherwise (e.g., in case of octave errors).}

From Eqn.~\eqref{eq:res}, if both conditions $\eqref{eq:cond}$ and $\eqref{eq:condzero}$ hold, then $\LL{equiv} = 0$. However, $\LL{equiv}$ can be null without these conditions to be satisfied. Hence, there is a need for a third loss term.

\subsubsection{Regularization loss}

Since we want $\bs{y}$ and $\bsk{y}$ to be identical up to a translation of $k$ bins, we propose to minimize the \ac{SCE} of one relative to the other.
In practice, we pad both distributions with zeros on both sides, i.e., for any integer $i \in \ZZ$, if $i < 1$ or $i > K$, then $\widetilde{y}_i = \sk{y}_i = 0$.

The \ac{SCE} loss $\LL{SCE}$ is then defined as
\begin{equation}
    \LL{SCE}(\bs{y}, \bsk{y}, k) = \sum_{i=1}^K \sk{y}_{i+k} \log \widetilde{y}_i.
\end{equation}

This loss is minimal if and only if the conditions from Eqn.~\eqref{eq:cond} and \eqref{eq:condzero} are met. However, contrary to $\LL{equiv}$, the (shifted) cross-entropy does not take into account the ordering of the probability bins, making the value of $\LL{SCE}$ independent of the actual pitch error between $\bs{y}$ and $\bsk{y}$. We, therefore, combine both losses.

\subsubsection{Loss weighting}\label{sec:gradients}

% \begin{itemize}
%     \item conceptually, we want our final objective to be symmetric
%     \item However, cross-entropy is not a symmetric distance but defined for one distribution relative to another one
%     \item in practice, we make our loss symmetric by swapping terms
%     \item in our preliminary experiments, we found that stopping the gradients of the target distributions stabilizes training. Our whole loss objective is therefore
% \end{itemize}

Conceptually, there is no reason for our final objective not to be symmetric. However, cross-entropy is not inherently symmetric as it is defined for one distribution relative to another; we therefore symmetrize it by swapping terms.
Additionally, we observed in our preliminary experiments that stopping the gradients of the target distributions helps stabilize training. Therefore, our complete loss objective incorporates this gradient-stopping mechanism \new{(denoted by $\sg$)} and is written as follows:
\begin{equation}
    \begin{aligned}
        \LL{}%(\b{y}, \bs{y}, \bsk{y}, k) 
        &= \frac{\lambda_\text{inv}}{2} \left( \LL{inv}(\b{y}, \sg (\bs{y})) + \LL{inv}(\bs{y}, \sg(\b{y})) \right) \\
        &+ \frac{\lambda_\text{equiv}}{2} \left( \LL{equiv}(\bs{y}, \bsk{y}, k) + \LL{equiv}(\bsk{y}, \bs{y}, -k) \right) \\
        &+ \frac{\lambda_\text{SCE}}{2} \left( \LL{SCE}(\bs{y}, \sg(\bsk{y}), k) + \LL{SCE}(\bsk{y}, \sg(\bs{y}), -k) \right)
    \end{aligned}
\end{equation}

The weights $\lambda_*$ are updated during training using the respective gradients of the losses concerning the last layer $\b{w}$ of the network $f_{\theta}$, following, e.g., \citet{VQGAN,MacGlashan2022}.

Let $S = \{ \text{inv}, \text{equiv}, \text{SCE} \}$. For $s \in S$, we first compute the following quantity:
\begin{equation}
    g_s = \dfrac{\left\| \nabla_{\b{w}} \mathcal{L}_s \right\|}{\sum_{s' \in S} \left\| \nabla_{\b{w}} \mathcal{L}_{s'} \right\|},
\end{equation}
where $\| \cdot \|$ stands for the $L_2$ norm.

Indeed, as studied in \citet{MacGlashan2022}, the norm of the gradient of each loss can be interpreted as its contribution to the total objective to optimize. To balance the contributions of each loss, we therefore weight each loss $\mathcal{L}_s$ with the sum of the contributions of the other losses, i.e., $1 - g_s$.

To prevent brutal variations of $\lambda_s$, in practice we update it as an \ac{EMA} of the $g_s$, i.e.,
\begin{equation}
    \begin{cases}
        \lambda_s(0) = 1 \\
        \lambda_s(t+1) = \eta \lambda_s(t) + (1 - \eta) (1 - g_s) \\
    \end{cases}.
\end{equation}

\subsection{Architecture}\label{sec:architecture}

The architecture of our pitch estimator is inspired by~\cite{Weiss2022, BasicPitch}.Each input VQT frame is processed independently through the following sequence: first, layer normalization ~\citep{LayerNorm} is applied, followed by a series of 1D convolutions along the log-frequency dimension - three with skip-connections ~\citep{ResNet} and four regular ones. The kernel size is set to \morenew{$13B = 39$} to span more than one octave \citep{BasicPitch} \morenew{- for $B=3$ bins per semitone}. As in \cite{Weiss2022}, we apply a non-linear leaky-ReLU with a slope of $0.3$ ~\citep{LeakyReLU} and dropout with rate $0.2$ ~\citep{Dropout} between each convolutional layer. 
Importantly, the kernel size and padding of each of these layers are chosen so that the frequency dimension is never reduced. A translation of $k$ bins of a \ac{VQT} frame, therefore, leads to a translation of $k$ bins of the output pitch distribution.
The output is then flattened, fed to a final fully connected layer, and normalized by a softmax layer to become a probability distribution of the desired shape.

Note that all layers (convolutions\footnote{Convolutions roughly preserve translations since the kernels are applied locally, meaning that if two translated inputs are convolved by the same kernel, then the output results will be almost translations of each other as well.}, elementwise non-linearities, layer-norm, and softmax), except the last final fully-connected layer, preserve transpositions.
To make the final fully connected layer also transposition-equivariant\new{, and in line with previous works on fully-convolutional networks~\citep{Long2014,FCNF0}}, we propose to use \textbf{Toeplitz fully connected layers}. It consists of a standard linear layer without bias but whose weights matrix $A \in \RR^{F' \times K}$ is a Toeplitz matrix, i.e., each of its diagonals is constant.

\begin{equation}
    \label{eq_toeplitz}
    A =
    \begin{pmatrix}
        a_0	    & a_{-1} & a_{-2} & \cdots & a_{-K+2} & a_{-K+1} \\
        a_1	    & a_0    & a_{-1} & \ddots & \ddots & a_{-K+2} \\
        a_2     & a_1    & \ddots & \ddots & \ddots & \vdots \\
        \vdots  & \ddots & \ddots & \ddots & \ddots	& \vdots \\
        a_{F'-1} & \cdots & \cdots & \cdots & \cdots	& a_{F'-K}
    \end{pmatrix}
\end{equation}

%Contrary to arbitrary, fully-connected layers, Toeplitz matrices are transposition-preserving operations with only $m + n - 1$ parameters instead of $mn$.
%Furthermore, they are mathematically equivalent to convolutions, making them straightforward to implement.
In practice, this matrix $A$ is equivalent to a 1D-convolution with kernel \morenew{$(a_{-K+1}, \dots, a_{F'-1})$ and padding $K - 1$}, and therefore preserves translations while having less parameters than an arbitrary fully-connected layer.

\subsection{Re-centering pitch distributions}

Recall that our model learns to predict pitches from VQT frames up to an additive constant and that the bins of the predictions do not correspond one-to-one to VQT log-frequencies: during training, the bin associated with a given pitch is completely arbitrary and only depends on the initialization of the model.

In practice, due to random weight initialization, we sometimes observed that the performances of our model significantly drop when a low pitch value is initially mapped to a high bin of the pitch distribution (or the opposite) because the whole pitch distribution is concentrated in a few bins. This is particularly likely when training on datasets that span a wide range of frequencies (such as MDB-stem-synth). While this phenomenon is far from the norm, it can limit the usability of our model in real-world scenarios.

To prevent that, we set our output probability distribution to cover more than $10$ octaves (\new{$K = 384$} bins with a resolution of $B = 3$ bins per semitone), making collapse unlikely for most real-world data.
In addition, we explicitly force the median of all the predictions to remain roughly in the center of the pitch distribution \new{during training}. If it deviates too much, we apply a circular shift to the kernel $(a_{m-1}, \dots, a_{-n+1})$ of the final Toeplitz fully-connected layer, which moves the predicted pitch distributions accordingly.\footnote{Precisely, we force the median of the pitch distribution to lie between bins 144 and 240 \new{at the end of each epoch}.} This simple trick enables us to rectify bad initialization during training with minimal overhead.

\begin{figure*}
    \centering
    \begin{subfigure}{0.49\textwidth}
        \includegraphics[width=\linewidth]{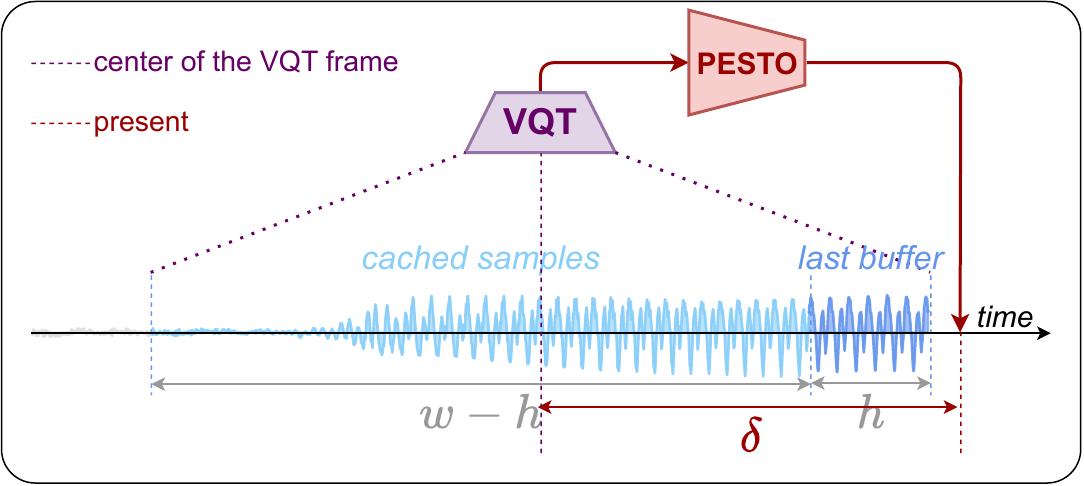}
        \caption{No buffer refilling ($m = 0$)}
        \label{fig:realtime-nomirror}
    \end{subfigure}
    \hspace*{\fill}
    \begin{subfigure}{0.49\textwidth}
        \includegraphics[width=\linewidth]{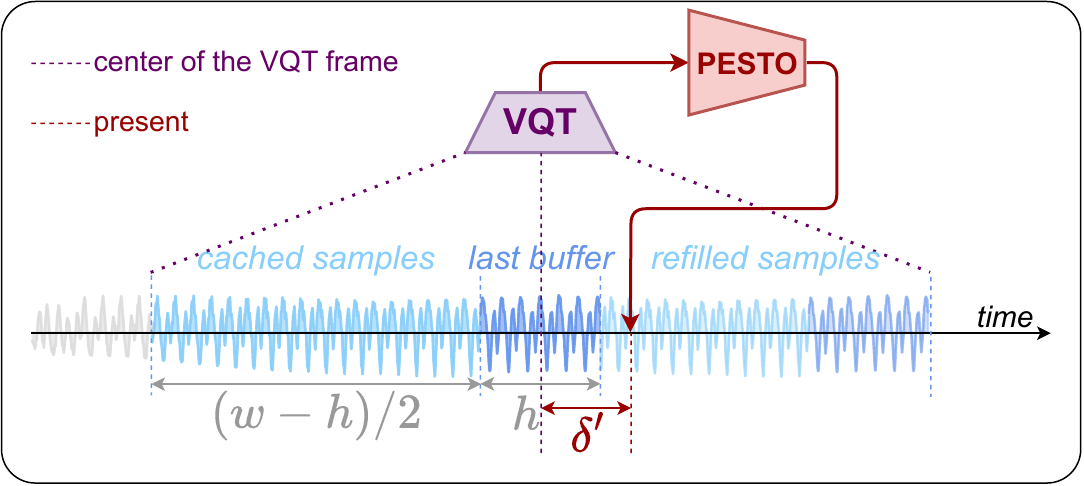}
        \caption{Full buffer refilling ($m = 0.5$)}
        \label{fig:realtime-mirror}
    \end{subfigure}
    \caption{Illustration of the latency of our model, and how to mitigate it with buffer refilling.
    %The computation time of our model is lower than the length $h$ of a buffer, which enables real-time applications. However, our model estimates the pitch of the center of the current VQT frame (purple dashed line), which is delayed by $w/2$ compared to the end of the current buffer.
    When a new buffer is consumed, the returned prediction is the pitch of the center of the VQT frame. Therefore, there is a delay of $w/2$ between when a buffer of audio is obtained and its actual pitch is estimated. Buffer refilling places the most recent buffer at the center of the processed VQT frame, thus improving the reactivity of the model.}
    \label{fig:realtime}
\end{figure*}

\subsection{Pitch decoding}

\subsubsection{Inferring pitch values from a pitch distribution}

For a given input VQT frame, let $\y = (y_1, \dots, y_K) \in \RR^K$ be the output of PESTO (a pitch distribution in the log-frequency domain). Relative pitch decoding is inferred by applying \emph{argmax-local weighted averaging} \citep{CREPE, PENN} to $\y$:

\begin{equation}
    p_\text{rel}(\y) = \dfrac{\sum_{i=a-2B}^{a+2B} i*y_i}{\sum_{i=a-2B}^{a+2B} y_i},
\end{equation}

\noindent where $a=\argmax(\y)$ and $B= 3$ is the number of bins per semitone in the VQT. Although in this work pitch decoding is only used for evaluation, we note that it can easily be made differentiable by using the expected value of the full pitch distribution \citep{DDSPinv, TORRES-ICASSP}.

%\TODO{make this equation less ugly}

\subsubsection{From relative to absolute pitch}

Conversion from relative ($p_\text{rel}$) to absolute MIDI pitch ($\hat{p}$) is performed  by applying the affine mapping:
\begin{equation}
    \hat{p}(\y) = \dfrac{1}{B} \left( p_\text{rel}(\y) + p_0 \right),
\end{equation}
\noindent where $p_0$ is a fixed shift that only depends on the trained model. As in \cite{SPICE}, we calibrate \new{this} shift $p_0$ by relying on a set of synthetic data with known pitch. We synthesize harmonic signals $\{s_{f_0=j}\}_{j=60}^{84}$ with $5$ harmonics, fundamental frequency values ranging from MIDI pitch $60$ to $84$ and harmonic amplitudes and overall gain drawn from a uniform distribution. $p_0$ is computed as the median distance between the predicted pitches and the ground-truth fundamental across all synthetic signals. 

\section{Real-time pitch estimation} \label{sec:real_time}

%\alain{In the end I think I prefer having real-time in a separate part}

Our model is very lightweight and achieves a processing speed significantly faster than real-time, which makes it theoretically well-suited for real-time applications (see Section \ref{sec:speed}).
However, practical real-time performance requires considerations beyond raw compute speed, as it requires the model to be causal and able to process individual audio streams.

In this section, we describe a streamlined approach for implementing real-time inference, allowing the model to process audio streams with minimal latency.
%Notably, our approach relies on post-training modifications of the original model, avoiding the need for retraining and thus enhancing its practical applicability.

\subsection{Streamable Variable-\textit{Q} Transform}

By design, our model processes individual VQT frames independently, which makes it inherently suitable for real-time applications. However, given an audio stream $\b{s} = (s_1, s_2, \dots)$, computing a VQT frame centered around $s_i$ requires frames from $s_{i-w}$ to $s_{i+w}$, where $w$ is the largest VQT window size. This non-causal operation is incompatible with real-time processing, as it requires future audio samples.

We build upon nnAudio \citep{nnAudio}, which implements audio time-frequency transforms as PyTorch modules using convolutional layers.\footnote{\url{https://github.com/KinWaiCheuk/nnAudio}} In nnAudio's time-domain implementation, the transforms basis kernels are precomputed and stored as frozen model parameters. While nnAudio computes the CQT by convolving the input audio with CQT kernels followed by normalization, we modify this approach by replacing the CQT kernels with VQT kernels. However, the convolution kernel lengths still exceed the typical buffer size of 5 to \SI{20}{\milli\second} needed for real-time applications (see \autoref{fig:vqt}). To address this, we replace standard convolutional layers with \emph{cached convolutions}, allowing our model to support streamed inputs by storing previous audio chunks in an internal circular buffer~\citep{CachedConv}.

\subsection{Lag minimization}\label{sec:lag}
% \alain{re-read that stuff, there's likely bullshit in it}
% Our model's prediction at time $t$ corresponds to the input audio's pitch at time step $t-\frac{w-h}{2f_s}$, where $w$ is the kernel size, $h$ is the hop size, and $f_s$ is the sampling rate. This introduces a theoretical lag $\delta$ between the audio input and the pitch prediction:

% \begin{equation}
%     \delta = \dfrac{w}{2 f_s},
% \end{equation}
Our model predicts the pitch of the center of the VQT frames given as input. In addition to the compute time $\tau$ of a forward pass of our model (typically less than \SI{10}{\milli\second}), this implies a theoretical lag equal to half of the VQT's kernel size between when a buffer is acquired and when the pitch of a frame centered on this buffer is returned:
\begin{equation}
    \delta = \dfrac{w}{2 f_s} + \tau
\end{equation}

\noindent where $w$ represents the VQT largest kernel size and $f_s$ is the sampling rate (see \autoref{fig:realtime-nomirror}). Given that $w$ is typically proportional to $f_s$ (see Eqn.~\eqref{eq:window}), the main limiting factor is the window size.

For the CQT, the kernel size $w$ is particularly large at low frequencies (e.g., $w = 131072$ samples, or $2.7$ seconds, at \SI{48}{\kilo\hertz} for an A0). By using the VQT, this issue is alleviated by reducing $w$ drastically, with the added bonus of enhancing model performance (see Section \ref{sec:ablation_frontend}). For instance, by selecting $\gamma = 7$, we have a maximum kernel size of $w = 8192$ samples at \SI{48}{\kilo\hertz} (171 ms). Even with such improvements, a lag of around \SI{70}{\milli\second} remains perceptible for most applications, such as real-time resynthesis.

%To further reduce latency, we experimented with a sliding VQT~\citep{SlidingCQT} approximation, though preliminary results indicated a detrimental effect on performance.
Increasing $\gamma$ could further decrease $w$ (see \autoref{fig:vqt}), but beyond a certain point, this also compromises model accuracy, as shown in Section \ref{sec:ablation_frontend}.

%\alain{not sure we need continuous mirroring values}
To minimize this lag, we propose a buffer refilling technique that artificially creates windows centered closer to the current buffer. 
By shifting the input buffer to the centre of the VQT frame, and filling the right side of the window (i.e. the non-causal region) by repeating samples from the end of the buffer, we are able to predict the pitch of the current audio chunk with minimal latency.
For a causal audio stream $\b{s} = (s_1, \dots, s_t)$, the frame is computed from a modified chunk:
% $$(s_{t - w + \lfloor m(w-h) \rfloor}, \dots, s_{t-1}, s_t, s_{t-1}, \dots, s_{t - \lfloor m(w-h) \rfloor + 1}) \in \mathbb{R}^w,$$
\begin{equation}
    (\underbrace{s_{t - w + \lfloor m(w-h) \rfloor}, \dots, s_{t-1}, s_t}_{\text{past samples}}, \underbrace{s_{t-\lfloor m(w-h) \rfloor}, \dots, s_t}_{\text{refilled samples}}) \in \mathbb{R}^w,
\end{equation}
where \new{$h$ is the hop/buffer size and} $m \in \left[ 0, \frac{1}{2} \right]$ is the refill factor, with $m = 0$ corresponding to no refilling at all (default behavior) and $m = \frac{1}{2}$ corresponding to the maximum possible refilling.

While it is common to reflect samples when repeating data as padding at analysis window boundaries, buffer refilling keeps the repeated samples in their original order.
This is because reflection in the middle of the window would cause destructive interference between the imaginary components of the CQT kernel, which is conjugate symmetric under reversal.

As illustrated in \autoref{fig:realtime-mirror}, this technique enables us to construct windows centered on the most recent audio buffer, leading to:
% \begin{equation}
%     \delta' = \dfrac{(1 - 2m)w}{2 f_s} + \tau.
% \end{equation}
\begin{equation}
    \delta' = \dfrac{1}{f_s}\left\lfloor\frac{1}{2}w - m(w-h)\right\rfloor + \tau, \quad m\in\left[0,\frac{1}{2}\right].
\end{equation}
In particular, the total lag can be reduced to $h /2f_s+\tau$ using maximum refilling ($m=0.5$), making the system extremely reactive.
%The impact of this mirroring technique is studied in Section~\ref{sec:mirroring_effect}, where we demonstrate its efficacy empirically. 

%In addition to this theoretical lag, there is also a practical one induced by the processing time between an audio chunk being received and the pitch prediction from our model being returned.
To reduce the overall latency of our model, we also aim at reducing the compute time $\tau$ of our model itself. We improve the original implementation of nnAudio's VQT by computing both real and imaginary kernels with a single convolution layer, and further reduce compute time with JIT compilation.

% On top of that, we also propose to replace the final Toeplitz fully-
% To reduce it, we propose several techniques, which we measure the efficacy in Table \TODO{not sure everything is relevant, maybe just talk about fftconv and compare conv to fftconv, I trust ben}
% \begin{itemize}
%     \item JIT compilation
%     \item fftconv from Ben
%     \item We also concatenate the real and imaginary parts of the VQT kernels to compute both parts in parallel with a single convolution instead of sequentially. \TODO{not sure if everything is very relevant to include in a paper, let's see}
% \end{itemize}

\subsection{Frequency-domain Toeplitz Layer}\label{sec:freq}

Further, we propose a faster implementation of the final Toeplitz fully-connected layer of our architecture (see section \ref{sec:architecture}).
This layer, necessary for PESTO's translation equivariance, can be further optimised to reduce runtime complexity.
Na{\"i}ve implementation of the layer as a one-dimensional convolution incurs a cost of $\mathcal{O}((m + n - 1) \cdot n)$.
\new{This can be reduced to $\mathcal{O}(m\cdot n)$ through optimisations such as implicit padding, as implemented in optimized backends such as oneDNN and cuDNN, or even by directly realizing the $m\times n$ Toeplitz matrix.}
However, given the size of both the activation vector and implicit kernel, time complexity can be further reduced by performing convolution as multiplication in the frequency domain.
\new{This allows the Toeplitz layer to be applied in log-linear time} and eliminates redundancy in the model's forward pass.

%\subsubsection{Influence of mirroring}

%\input{tables/right-pad}

\section{Experimental setup} \label{sec:experimental_setup}

\subsection{Implementation details}\label{sec:implementation}

\new{Unless specified otherwise, VQT computations are performed with $f_{\min} = 27.5$ Hz, which is the frequency of the lowest key of the piano (A0), $B=3$ bins per semitones and at most $F = 99B$ log-frequency bins, which corresponds to the maximal number of bins respecting the Nyquist frequency for a $16$ kHz signal. The maximal pitch-shift between frames is $k_{\max} = 16$ bins, i.e. slightly more than $5$ semitones.}
We set the VQT parameter $\gamma=7$ for our main experiments \new{and use a resolution of $B = 3$ bins per semitone}.

\new{For our equivariance loss, we fix $\alpha = 2^{\frac{1}{12B}}$. The EMA rate for the loss weighting is $\eta = 0.999$.}
White noise and random gain are applied as pitch-preserving augmentations with a probability of $0.7$ to the cropped VQT frames.
Standard deviation values for the white noise are drawn from a uniform distribution between $0.1$ and $2$, and gain values are drawn between $-6$ and $3$ dB. For training, we use a batch size of $256$ and the Adam optimizer \citep{Adam} with a learning rate of $10^{-4}$ and default parameters. The model is trained for $50$ epochs using a cosine annealing learning rate scheduler.
\new{We refer to our code for precise configurations and hyperparameters.}
Our architecture being very lightweight, training requires only 845MB of GPU memory and can be performed on a single GTX 1080Ti.

\subsection{Datasets}

Three datasets are considered for this study:

\begin{enumerate}
    \item \textbf{MIR-1K} \citep{MIR-1K} contains $1000$ tracks (about $2$h)
    of amateur singing of Chinese pop songs, with separate
    vocal and background music tracks provided. \new{The isolated vocals partition is used for most experiments, apart from the one described in Section \ref{sec:robustness_ablation}.}
    
    \item \textbf{MDB-stem-synth} \citep{MDBstemsynth} contains 15.56 hours of re-synthesized monophonic music played by 25 different instruments. Perfect fundamental frequency annotations are available.
    \item \textbf{PTDB} \citep{pirkerPitchTrackingCorpus2011} contains 4718 speech recordings from English speakers speech and corresponding laryngograph recordings for a total of $9.6$ hours.
\end{enumerate}

The datasets vary in \new{size,} pitch range and granularity for the provided ground-truth pitch annotations. Pitch annotations are provided with a hop size of \SI{20}{\milli\second} for MIR-1K, \SI{2.904}{\milli\second} ($128$/$44100$) for MDB-stem-synth and \SI{10}{\milli\second} for PTDB. We opt not to resample the annotations, as our model is not bound to a fixed hop size or sample rate. During training, we can take advantage of the original granularity of the annotations, while on inference, we can choose the hop size that best fits the application. This is another advantage of using SSL with time-frequency frontends, compared to supervised learning and raw waveform training: the sample rate and hop size can be changed on the fly for inference. 

As noted in previous work \citep{PENN}, a crucial problem in neural pitch estimation is overfitting to the pitch characteristics in the training data. Therefore, we conduct separate model training and evaluation on all datasets and examine generalization performance through cross-evaluation. Another practical issue is the potential misalignment of pitch annotations, that can be difficult to detect and potentially worsen the performances of a model trained on these.
\new{For instance, on PTDB pitch annotations are provided at time steps with a $5$ms offset (e.g., $5$ms, $15$ms, $25$ms for a hop size of $10$ms) compared to MIR-1K/MDB-stem-synth, where annotations are aligned with multiples of the hop size (e.g., $0$ms, $10$ms, $20$ms). Indeed, we observed that not accounting for this offset \morenew{negatively impacts} the evaluation metrics.}
Since our model does not use annotations at training time, it is not affected by this issue and potential misalignments can be accounted for without retraining.

\subsection{Baselines}

We compare our model to state-of-the-art SSL and supervised neural pitch estimators:
\begin{itemize}
\item \textbf{CREPE} \citep{CREPE} is a supervised model trained on MDB-stem-synth \citep{MDBstemsynth} and MIR-1K \citep{MIR-1K}, as well as Bach10 \citep{Bach10}, RWC-Synth \citep{pYIN}, MedleyDB~\citep{MedleyDB} and NSynth~\citep{NSynth}. We use the \new{default pretrained model provided in the official repository and deactivate Viterbi smoothing for fair comparisons}.\footnote{\url{https://github.com/marl/crepe}}
\item \textbf{PENN} \citep{PENN} is a supervised model which builds upon the FCNF0 architecture by implementing several and training improvements. We do not retrain PENN, but we reproduce the authors' results on MDB-stem-synth and PTDB with the publicly available \texttt{FCNF0++} model \footnote{\url{https://github.com/interactiveaudiolab/penn}}, which was trained on the training partitions of both datasets.
\item \textbf{SPICE} \citep{SPICE} is an SSL model trained to minimize the pitch difference (up to the known shift) between two cropped CQT frames. SPICE additionally reconstructs the input frame using a decoder and employs a reconstruction regularization.
\item \textbf{DDSP-inv} \citep{DDSPinv} performs self-supervised pitch estimation by using analysis-by-synthesis and estimating the parameters of a harmonics plus noise synthesizer, with pretraining on synthetic data.
\end{itemize}

\subsection{Evaluation metrics}

Our evaluation procedure is standard and common to previous works~\citep{CREPE,SPICE,PENN}. In practice, estimated and reference frequencies are converted to fractional semitones using the mapping:
\begin{equation}
    \texttt{hz2mid} : f \mapsto 12 \log_2 \left( \frac{f}{440} \right) + 69
\end{equation}
which maps the A4 (\SI{440}{\hertz}) to $69$ (which is the MIDI standard) and the other pitches accordingly.%\footnote{Some papers use cents instead of fractional semitones, however this is equivalent since 1 semitone = 100 cents.}

For any voiced frame, the \emph{pitch error} $e$ between the estimated fundamental frequency $\hat{f}$ and the ground truth $f$ is therefore:
\begin{equation}
    \begin{aligned}
        e(\hat{f}, f)
        &= \left| \texttt{hz2mid}(\hat{f}) - \texttt{hz2mid}(f) \right| \\
        &= 12 \left| \log_2 \left( \hat{f} / f \right) \right|
    \end{aligned}
\end{equation}
which exactly corresponds to the (fractional) number of semitones between the prediction and ground truth pitch. From this error, we compute the standard metrics introduced by \citet{Poliner2007}:
\begin{itemize}
    \item \textbf{\ac{RPA}}, which measures the proportion of voiced frames for which the pitch error is less than half a semitone ($50$ cents).
    \item \textbf{\ac{RCA}}, which measures the proportion of voiced frames for which $|e(\hat{f}, f) \mod 12| \leq 0.5$, thus accepting octave errors.
\end{itemize}
In addition, we also report the \ac{MnE} and \ac{MdE} \new{in semitones} over all voiced frames.

\section{Results}\label{sec:results}

\input{tables/results-ssl}

\input{tables/results_color}

\subsection{Comparison with self-supervised methods}

PESTO significantly advances the state-of-the-art in self-supervised pitch estimation, achieving  $97.7\%$  and $97.0\%$ Raw Pitch Accuracy (RPA), on MIR-1K and MDB-stem-synth respectively, outperforming previous self-supervised approaches SPICE and DDSP-inv by a large margin
(Table \ref{tab:results-ssl}).

\subsection{Comparison with supervised methods}
Table \ref{tab:results} shows experimental results comparing PESTO with supervised methods. Despite having only $130$k parameters—$170$ times fewer than CREPE's $22.2$M—our model exceeds its performance on MIR-1K ($97.7\%$ vs $97.5\%$ RPA) and PTDB (\morenew{$89.7\%$} vs $87.1\%$ RPA), while being comparable on MDB ($97\%$ vs $97.3\%$).
 While PENN achieves superior results when trained on matched domains ($99.6\%$ RPA on MDB, $95.1\%$ on PTDB), PESTO maintains competitive performance without requiring supervision.

%\TODO{decide which numbers to put here}

\subsection{Cross-dataset evaluation}

Table \ref{tab:results} shows detailed results for cross-dataset evaluation.
\stefan{Also discuss where PENN performs better than Pesto (e.g. MDB stem synth, PTDB. Generally, discuss objectively, don't only mention where PESTO performs better.))}\alain{in the cross-dataset scenario, PESTO always performs significantly better than PENN. That's not true in other scenarios and I emphasized this more in }
\medskip

\noindent \textbf{Music  $\to$ Speech}:  PESTO models trained on music data outperform CREPE and PENN (when trained on music) on speech datasets. PENN, when trained only on MDB, achieves $63.2$ RPA on PTDB, while PESTO achieves $88.3$. CREPE, trained on many more music datasets, achieves $87.1$ RPA on PTDB.

\medskip

\noindent \textbf{Speech $\to$ Music}:  When trained on PTDB (speech), PESTO still achieves $96.3\%$ RPA on MDB (music), while PENN's performance drops to $51.6\%$. 

\medskip
\noindent \textbf{Music/Speech  $\to$  Singing Voice}: PESTO consistently outperforms PENN on MIR-1K across training conditions.  When trained on both MDB and PTDB, PENN achieves  $90.6\%$ RPA, while PESTO achieves $94.6\%$ RPA or higher in all cross-evaluations for MIR-1K.

\medskip

\noindent Overall, our results corroborate the findings that supervised models are heavily influenced by the training pitch distribution \citep{PENN} and show that PESTO is more robust to out-of-distribution data. Although PESTO training and evaluating on the same dataset yields the best results, the cross-domain performance gap remains considerably smaller than supervised baselines. Notably, the MdE metrics remain stable across training sets, suggesting PESTO captures fundamental pitch characteristics while dataset-specific training primarily improves robustness to challenging frames. 

\subsection{Multi-Dataset Training}

When evaluating on a given dataset, we observe that models trained solely on that dataset generally have better performance than models trained on combined datasets. MDB is a notable exception where comprehensive training yields the best results, though the difference is marginal ($97.1\%$ RPA instead of $97.0\%$).
This effect may be attributed to MDB's larger size dominating the combined dataset distribution.

On the contrary, PENN performs better (on PTDB) or equally (on MDB) when trained on multiple datasets, and clearly outperforms PESTO when the training and test distributions overlap.

\input{tables/speed_benchmark}

\subsection{Inference speed}\label{sec:speed}

We evaluated the real-time capabilities of our PESTO model,
%(in both streaming and non-streaming modes)\alain{we probably don't need streaming here)
as well as PENN \citep{PENN} and YIN \citep{YIN} as baselines, without Viterbi decoding. The Real-Time Factor (RTF), defined as the ratio of processing time to audio input duration, was measured across CPU (Intel Core i9-12900H 2.9GHz) and GPU (NVIDIA RTX A2000 8GB) platforms. Using 5-second audio segments at \SI{16}{\kilo\hertz} with a \SI{10}{\milli\second} hopsize, we performed 100 measurements for each configuration. For PENN GPU inference, we used a batch size of 2048 frames using the publicly available API. All reported timings in Table \ref{tab:rtf_comparison} include the inference pipeline from loaded audio input to pitch estimation (or pitch candidates for YIN) output.

% \subsection{Failure cases}

% - 

PESTO outperforms all other models in RTF by a significant margin. PENN achieves real-time performance (in domain) but requires roughly 5 times more processing time than PESTO on CPU and 3 times on GPU,  while using parallel computation of frames with a batch size of $2048$. 
%The YIN baseline is approximately $60\%$ slower than PESTO on CPU and 18 times times slower on GPU.
Finally, PESTO is approximately $60\%$ faster than the YIN baseline on CPU and 18 times faster when ran on a GPU.

\input{tables/mirror}

\subsection{Effect of buffer refilling}

To determine the effect of the lossy buffer refilling technique (Section \ref{sec:lag}) on PESTO's performance, we retrained the model for refill factors $m\in\left\{0.1,0.2,\dots,0.5\right\}$ on the MIR-1k dataset.
RPA and RCA metrics are reported on the test dataset in Table~\ref{tab:mirror}. 
We observe a minimal degradation in performance, with a RPA of $97.4\%$ at the maximum possible refill factor ($m=0.5$), compared to the standard PESTO model which achieves $97.7\%$ ($m=0.0$).
This suggests that PESTO is relatively robust to the removal of future context via buffer refilling, and is thus suitable for deployment in real-time and latency critical scenarios.

\new{For comparison, we also try replacing the refilled portion of the analysis window with zeros.
For $m \leq 0.4$ this appears to perform equivalently to buffer refilling, but at the maximum refill factor of $m=0.5$ we find that the zero-filling technique suffers degraded performance.}

\subsection{Robustness to background music} \label{sec:robustness_ablation}

In real-world scenarios, we often do not have access to clean signals. In this section, we evaluate to what extent the predictions of our model are robust to the presence of background music in the signals.
To do so, we use the MIR-1k dataset, for which we have access to the separated vocals and background music, which allows testing various signal-to-noise (here vocals-to-background) ratios.

We indicate the results in \autoref{tab:background}. As foreseen, the RPA of PESTO when trained only on clean vocals (row $\beta = 0$) considerably drops: from $97.2$\% to $46.8$\%.

To improve the robustness to background music, we propose to modify the training pipeline slightly. Instead of the data augmentations described in Section~\ref{sec:implementation}, we create the augmented view $\bs{x}$ of our original vocals signal $\x_{\text{vocals}}$ by mixing it (in the complex-CQT domain) with its corresponding background track $\x_{\text{background}}$:
\begin{equation}
    \bs{x} = \dfrac{1}{1 + |\beta|} \left( \x_{\text{vocals}} + |\beta| \x_{\text{background}} \right),
\end{equation}
with $\beta \in \RR$.\footnote{$\bsk{x}$ is obtained from $\bk{x}$ similarly, however the value $\beta$ is not necessarily identical.} The model is thus trained to ignore the background music to make its predictions.

We investigate different strategies for sampling $\beta$, whose respective results are depicted on \autoref{tab:background}.

\input{tables/background-rpa}
% \begin{itemize}
%     \item First, we observe that using background music as data augmentation always help whatever the sampling strategy.
%     \item in particular, sampling from a distribution instead of using a fixed value seems to work better
%     \item first because the model learns to cover different SNR
%     \item moreover because then the background level is different in $\bs{x}$ and $\bsk{x}$, so that not only $\LL{inv}$ contributes to ignore the background, but also the other losses
%     \item In particular, sampling $\beta$ from the uniform distribution $\mathcal{U}(0, \frac{1}{2})$ leads to the best performances in most scenarios, apart when the vocals and background are equally mixed ($\SNR = \SI{0}{dB}$), for which $\beta \sim \mathcal{N}(0, 1)$ seems to be optimal (82.5\% RPA).
%     \item Interestingly, using background music during training not only improves performances on noisy signals but also on clean ones: our best model reaches 98.3\% RPA on the clean test set of MIR-1K, which is higher than what we get with the previous data augmentations (see \autoref{tab:results}). This shows that, when available, background music can be very beneficial to single pitch estimators.
% \end{itemize}

\begin{figure*}
    \centering
    \includegraphics[width=\linewidth]{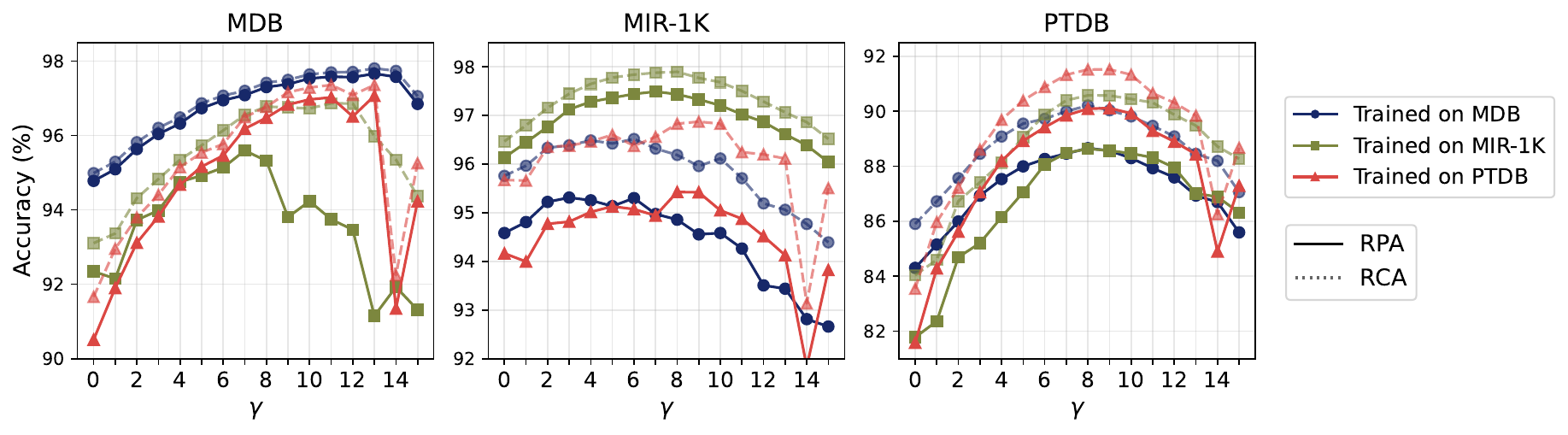}
    \caption{Comparison of pitch accuracy metrics across different datasets as a function of the VQT parameter $\gamma$. Each subplot shows test performance on a specific dataset (MDB, MIR-1K, or PTDB), with line colors and markers indicating the training dataset. \morenew{Solid} lines represent RPA, while \morenew{dashed} lines represent RCA. The points indicate the mean of the top 3 scores out of 5 runs with different random seeds.}
    \label{fig:gamma_sweep}
\end{figure*}

We observe that using background music as data augmentation consistently improves performance regardless of the sampling strategy. Notably, sampling from a distribution instead of using a fixed value yields better results. This improvement arises because the model learns to handle varying signal-to-noise ratios (SNR) and because the background level differs between the input $\bs{x}$ and its pitch-shifted counterpart $\bsk{x}$. Consequently, not only does the invariance loss $\LL{inv}$ help the model ignore the background, but the other losses also contribute.
Sampling $\beta$ from the uniform distribution $\mathcal{U}(0, \frac{1}{2})$ generally leads to the best performance, except when vocals and background are equally mixed ($\SNR = \SI{0}{dB}$), where $\beta \sim \mathcal{N}(0, 1)$ appears optimal, achieving $82.5$\% RPA. Interestingly, incorporating background music during training not only enhances performance on noisy signals but also improves results on clean ones. Our best model achieves $98.3$\% RPA on the clean test set of MIR-1K, surpassing previous data augmentation strategies (see \autoref{tab:results}). This demonstrates that when background music is available, it can significantly benefit single pitch estimators.

Finally, PESTO outperforms both supervised (CREPE, PENN) and self-supervised baselines (SPICE), apart \morenew{from} $\SNR = \SI{0}{dB}$ where CREPE performs better. However, the drop in performances between $\SNR = \SI{10}{dB}$ and $\SNR = \SI{0}{dB}$ is higher for PESTO (approximately $15$\%) than for SPICE or CREPE (about $10$\%), which suggests there is still room for improvement.

\begin{figure}[h!]
    \centering
    \includegraphics[trim={8 7 7 7},clip, width=0.95\linewidth]{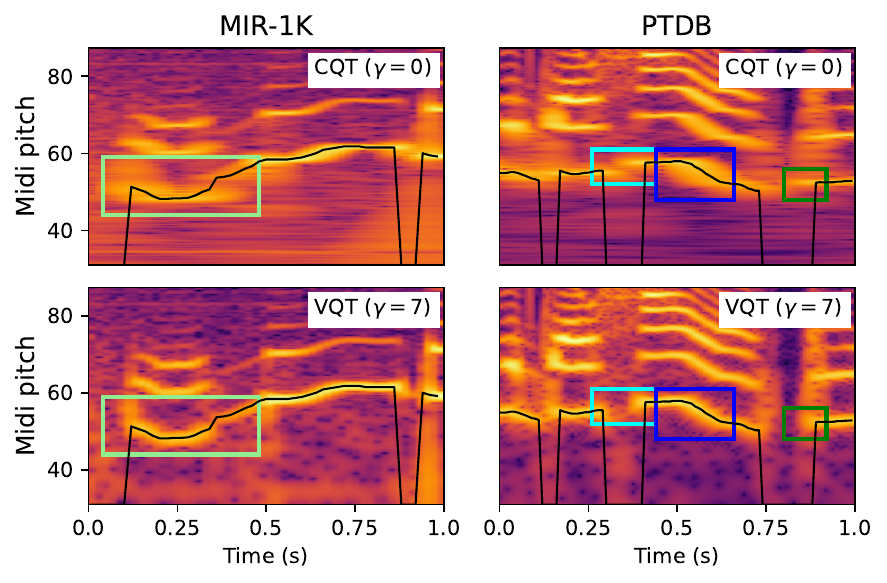}
    \caption{Comparison between CQT ($\gamma=0$, top) and VQT ($\gamma=7$, bottom) spectrograms for one example from MIR-1K (left) and PTDB-TUG (right) datasets. Highlighted rectangles indicate time/frequency misalignments between the spectrograms and the ground truth pitch contour (black line), particularly noticeable for the CQT at lower frequencies.}
    \label{fig:vqt_cqt}
\end{figure}

\section{Ablations}

\subsection{Impact of the input frontend}\label{sec:ablation_frontend}

We vary the $\gamma$ parameter of the VQT between $0$ (CQT) and $15$ and train models for each dataset. \autoref{fig:gamma_sweep} shows the RPA and RCA results for cross-evaluation for different values of  $\gamma$. Each reported value is the mean of the top $3$ scores achieved out of $5$ runs with different random seeds, for a given training dataset and $\gamma$ value. 

Our experiments demonstrate that the Variable-Q Transform (VQT) as an input to the model significantly enhances pitch estimation accuracy compared to the Constant-Q Transform (CQT). As it uses smaller kernel sizes at lower frequencies ($\gamma > 0$) than the CQT, the VQT achieves superior time-frequency alignment with the ground truth pitch contours. As an example, we illustrate in \autoref{fig:vqt_cqt} a comparison of spectrogram frames for the CQT ($\gamma=0$) and the VQT ($\gamma = 7$). The VQT  produces sharper temporal boundaries and more precise frequency localization than the CQT. 

While increasing $\gamma$ improves performance across all datasets, this benefit plateaus and eventually deteriorates when frequency resolution is compromised. This trade-off is particularly pronounced in the speech dataset, which has rapid phonetic transitions compared to musical signals yet still requires sufficient spectral detail for accurate pitch tracking. Interestingly, the optimal $\gamma$ value is not the same for all datasets, highlighting the fact that there is a domain-dependent tradeoff between low-frequency resolution and time resolution. Values in the range $\sim [7-10]$, however, provide the best results with relatively low variation for all datasets.

\input{tables/ablation}

In theory, under supervised training, one could leverage the temporal context to compensate for misalignments or learn the optimal time-frequency representation for pitch estimation. Our results suggest, however, that a small adjustment to the CQT already provides a more flexible and efficient solution that generalizes across diverse acoustic domains while allowing for SSL training without the need for labeled data. Furthermore, as Section \ref{sec:real_time} details, the use of the VQT significantly improves the computation speed of the model, which is critical for real-time.

\subsection{Design choices}

In this section, we examine the impact of various design choices on the performance of our model. The results for training on MIR-1K are shown in \autoref{tab:ablation_mir1k}. First, we observe that the final Toeplitz fully-connected layer, which ensures the architecture is transposition-preserving, is essential for the model to learn effectively. In contrast, data augmentations are not necessary for training but \morenew{usually} enhance performance.

The influence of the different loss functions is more complex to interpret. When using only $\LL{SCE}$, the model does not learn at all, whereas $\LL{equiv}$ alone yields good in-domain performance. However, combining $\LL{equiv}$ with the other two losses helps the model generalize to out-of-domain distributions. Interestingly, $\LL{inv}$ alone produces decent performance, indicating that learning to be insensitive to white noise and gain changes enables the model to focus on the fundamental frequency of the signal, though the accuracy is lower than when $\LL{equiv}$ is included.

Surprisingly, discarding $\LL{equiv}$ and using only the other two losses achieves the best out-of-distribution performance. This counterintuitive result prompted further evaluation of $\LL{equiv}$ when training on MDB-stem-synth (\autoref{tab:ablation_mdb}) and PTDB (\autoref{tab:ablation_ptdb}). These experiments, however, did not reveal a consistent pattern: $\LL{equiv}$ appears beneficial for training on PTDB but slightly detrimental for MDB-stem-synth.

This suggests that the design of our loss functions and the weighting strategy (based on gradient norms, see section~\ref{sec:gradients}) may benefit from further refinement.
%\subsection{Architecture}

\section{Conclusion}

% \begin{itemize}
%     \item In this paper, we introduce a novel self-supervised method for pitch estimation
%     \item We evaluate on several music and speech datasets and show it significantly outperforms previous self-supervised baselines, and is on par with supervised ones while having only 130k parameters
%     \item In particular, our method exhibits superior generalization capabilities than baselines, outperforming them when there is a shift between the training and test distribution
%     \item In addition, we propose a very simple solution to make our model able to process audio streams with arbitrary sampling rates or buffer sizes without requiring any retraining.
%     \item Combined with its very low latency (less than \SI{5}{\milli\second}), this makes it particularly suited for real-time applications.
%     \item Finally, the self-supervised paradigm we propose makes no musical assumption and is compatible with scenarios with very scarce annotated data, enhancing its applicability to non-Western music information retrieval~\citep{citeSthHere}, but also to other fields beyond \ac{MIR} such as bioacoustics~\citep{TODO} or even physics modelling~\citep{SoundOfWater}.
%     \item More generally, we strongly believe that the versatility of our method makes it suitable for a wide range of real-world applications.
%     \item With this in mind, we release our code and pretrained models in a pip-installable package, as well as the full training code, hoping that it will encourage its usage and further experimentation.
% \end{itemize}

In this paper, we introduced a novel self-supervised method for pitch estimation. Through evaluation on several music and speech datasets, we demonstrated that our method significantly outperforms previous self-supervised baselines and achieves performance on par with supervised approaches, despite having only $130$k parameters.

Our method exhibits superior generalization capabilities compared to baselines, particularly in scenarios where there is a shift between the training and test distributions. Additionally, we proposed a simple solution to enable the model to process audio streams with arbitrary sampling rates or buffer sizes, without requiring retraining. With its very low latency of less than \SI{5}{\milli\second}, this makes our approach highly suitable for real-time applications.

The self-supervised paradigm we propose is free from specific musical assumptions, making it applicable to tasks with very limited annotated data. This enhances its relevance for non-Western music information retrieval~\citep{Li2022Guzheng,Han2023Tori}, as well as for broader applications for which collecting annotated data can be impractical, such as bioacoustics~\citep{Best2022,ISPA} and physics modeling~\citep{SoundOfWater}.

The versatility of our method makes it suitable for a wide range of real-world applications. To encourage its adoption and further experimentation, we released our code and pretrained models as a pip-installable package, along with the full training pipeline.

Furthermore, exploiting equivariance for solving classification problems is a promising direction for future research, as it allows models to directly return probability distributions using only a single (potentially synthetic) labeled element. This has potential applications beyond pitch estimation, such as tempo estimation~\citep{Gagnere2024} or key detection~\citep{kongSTONESelfsupervisedTonality2024}.
%Additionally, the equivariance loss offers a differentiable way to convert a distribution into a scalar, which could have potential uses in differentiable digital signal processing (DDSP).~\TODO{not sure about this last sentence}

Finally, while our model focuses on monophonic pitch estimation, the training objective does not constrain it to a single prediction. This differs from previous self-supervised pitch estimation methods that frame the task as a regression problem~\citep{SPICE,DDSPinv}. In particular, it paves the way towards self-supervised multi-pitch estimation.

%%%%%%%%%%%%%%%%%%%%%%%%%%%%%%%%%%%%%%%%%%%%%%%%%%%%%%%%%%%%%%%%%%%%%%%%%%%%%%%%
% Please do not touch.
% Print Endnotes
\IfFileExists{\jobname.ent}{
   \theendnotes
}{
   %no endnotes
}
%%%%%%%%%%%%%%%%%%%%%%%%%%%%%%%%%%%%%%%%%%%%%%%%%%%%%%%%%%%%%%%%%%%%%%%%%%%%%%%%

\section*{Acknowledgments}

This work has been funded by the ANRT CIFRE convention n°2021/1537 and Sony France.
%This work was granted access to the HPC/AI resources of IDRIS under the allocation 2022-AD011013842 made by GENCI.

B. Torres and G. Richard are supported by the European Union (ERC, HI-Audio, 101052978). Views and opinions expressed are however those of the author(s) only and do not necessarily reflect those of the European Union or the European Research Council. Neither the European Union nor the granting authority can be held responsible for them.\\

We first thank the anonymous reviewers for their meticulous and constructive feedback, which greatly helped us improve the quality of this paper.
Moreover, we would like to thank Alexis André for motivating the need for a real real-time implementation of PESTO, and for pushing the boundaries of its artistic applications with his crazy generated visuals. We also thank Cyran Aouameur and Yanis Amedjkane for implementing a PESTO plugin in Juce, and for their insightful suggestions, as well as DeLaurentis for integrating PESTO on stage within her live performances. Finally, we thank Emmanuel Deruty for very inspiring discussions about the nature of pitch.

\section*{Author affiliations}

\noindent\textbf{Alain Riou}\\
\hangindent=1.5em \hspace{1.5em}LTCI, Télécom-Paris, Institut Polytechnique de Paris, France\\
\hangindent=1.5em \hspace{1.5em}Sony Computer Science Laboratories - Paris, France\\

\noindent\textbf{Bernardo Torres}\\
\hangindent=1.5em \hspace{1.5em}LTCI, Télécom-Paris, Institut Polytechnique de Paris, France\\

\noindent\textbf{Ben Hayes}\\
\hangindent=1.5em \hspace{1.5em}Centre for Digital Music, Queen Mary University of London, United Kingdom\\

\noindent\textbf{Stefan Lattner}\\
\hangindent=1.5em \hspace{1.5em}Sony Computer Science Laboratories - Paris, France\\

\noindent\textbf{Gaëtan Hadjeres}\\
\hangindent=1.5em \hspace{1.5em}Sony AI, Zurich, Switzerland\\

\noindent\textbf{Gaël Richard}\\
\hangindent=1.5em \hspace{1.5em}LTCI, Télécom-Paris, Institut Polytechnique de Paris, France\\

\noindent\textbf{Geoffroy Peeters}\\
\hangindent=1.5em \hspace{1.5em}LTCI, Télécom-Paris, Institut Polytechnique de Paris, France\\

%%%%%%%%%%%%%%%%%%%%%%%%%%%%%%%%%%%%%%%%%%%%%%%%%%%%%%%%%%%%%%%%%%%%%%%%%%%%%%%%
% Bibliography
%%%%%%%%%%%%%%%%%%%%%%%%%%%%%%%%%%%%%%%%%%%%%%%%%%%%%%%%%%%%%%%%%%%%%%%%%%%%%%%%

% For bibtex users:
\bibliography{macros, library,library2,ben, manual_bib_donot_ovewrite}

% For non bibtex users:
%\begin{thebibliography}{citations}
%
%\bibitem {Author:00}
%E. Author.
%``The Title of the Conference Paper,''
%{\it Proceedings of the International Symposium
%on Music Information Retrieval}, pp.~000--111, 2000.
%
%\bibitem{Someone:10}
%A. Someone, B. Someone, and C. Someone.
%``The Title of the Journal Paper,''
%{\it Journal of New Music Research},
%Vol.~A, No.~B, pp.~111--222, 2010.
%
%\bibitem{Someone:04} X. Someone and Y. Someone. {\it Title of the Book},
%    Editorial Acme, Porto, 2012.
%
%\end{thebibliography}

\end{document}

%% file: tables/results-ssl.tex
\begin{table}
    \centering
    \caption{Performances of our model compared to previous self-supervised methods. For both baselines, we report the results provided in the paper.
    %For DDSP-inv, we could not find a working implementation and therefore do not know the number of parameters of the model.
    }
    \label{tab:results-ssl}
    \small{
    \begin{tabular}{lccccc}
        \toprule
         &  &  & \multicolumn{2}{c}{Raw Pitch Accuracy} \\
        \cmidrule{4-5}
        Model & \# params &  & MIR-1K & MDB-stem-synth \\
        \midrule
        SPICE & 2.38M &  & 90.6 & 89.1 \\
        DDSP-inv & 21.6M &  & 91.8 & 88.5 \\
        PESTO & 130k &  & \textbf{97.7} & \textbf{97.0} \\
        \bottomrule
    \end{tabular}}
\end{table}

%% file: tables/results_color.tex
\begin{table*}
\definecolor{sameset}{rgb}{0.9, 0.9, 1.0}    % Light blue for same dataset
    \definecolor{crossset}{rgb}{1.0, 0.9, 0.9}    % Light red for cross-dataset
    \definecolor{multidataset}{rgb}{0.9, 1.0, 0.9}    % Light green for multi-dataset
    \centering
    \renewcommand{\thefootnote}{\fnsymbol{footnote}}
   \caption{Performance comparison across different datasets and training configurations. PESTO is self-supervised while CREPE and PENN are state-of-the-art supervised models. Colors indicate evaluation scenarios: \colorbox{sameset}{same-dataset}, \colorbox{multidataset}{multi-dataset}, and \colorbox{crossset}{cross-dataset} evaluation. For Raw Pitch Accuracy (RPA) and Raw Chroma Accuracy (RCA), higher is better, while for Mean Pitch Error (MnE) and Median Pitch Error (MdE) lower is better. Best results for each metric/test dataset are in bold.}
    \label{tab:results}
    \fontsize{7.3pt}{7.3pt}\selectfont
    
    \begin{tabular}{lccccccccccccccccccc}
        \toprule
         &  &  &  & \multicolumn{4}{c}{MIR-1K} &  & \multicolumn{4}{c}{MDB-stem-synth} &  & \multicolumn{4}{c}{PTDB} \\
        \cmidrule{5-8} \cmidrule{10-13} \cmidrule{15-18}
            Model & \# params & Training data &  & RPA & RCA & MnE & MdE &  & RPA & RCA & MnE & MdE &  & RPA & RCA & MnE & MdE \\
        \midrule
        CREPE & 22.2M & many &  &
        \cellcolor{multidataset}97.5 & \cellcolor{multidataset}\textbf{98.0} & \cellcolor{multidataset}0.23 & \cellcolor{multidataset}\textbf{0.07} &  & \cellcolor{multidataset}97.3 & \cellcolor{multidataset}97.4 & \cellcolor{multidataset}0.13 & \cellcolor{multidataset}\textbf{0.02} &  & \cellcolor{crossset}87.1 & \cellcolor{crossset}89.9 & \cellcolor{crossset}1.24 & \cellcolor{crossset}0.11 \\
        \midrule
        \multirow{3}{*}{PENN} & \multirow{3}{*}{8.9M} & MDB-ss &  &
         - & - & - & - &  & \cellcolor{sameset}99.7\footnotemark[2] & - & - & - &  & \cellcolor{crossset}63.2\footnotemark[2] & - & - & - \\
         &  & PTDB &  &
         - & - & - & - &  & \cellcolor{crossset}51.6\footnotemark[2] & - & - & - &  & \cellcolor{sameset}94.4\footnotemark[2] & - & - & - \\
         &  & both &  &
        \cellcolor{crossset}90.6 & \cellcolor{crossset}92.4 & \cellcolor{crossset}1.27 & \cellcolor{crossset}0.11 &  & \cellcolor{multidataset}\textbf{99.6} & \cellcolor{multidataset}\textbf{99.6} & \cellcolor{multidataset}\textbf{0.05} & \cellcolor{multidataset}0.03 &  & \cellcolor{multidataset}\textbf{95.1} & \cellcolor{multidataset}\textbf{96.0} & \cellcolor{multidataset}\textbf{0.30} & \cellcolor{multidataset}\textbf{0.06} \\
        \midrule
        \multirow{4}{*}{PESTO} & \multirow{4}{*}{130k} & MIR-1K &  &
        \cellcolor{sameset}\textbf{97.7} & \cellcolor{sameset}\textbf{98.0} & \cellcolor{sameset}\textbf{0.21} & \cellcolor{sameset}0.12 &  & \cellcolor{crossset}94.8 & \cellcolor{crossset}95.9 & \cellcolor{crossset}0.43 & \cellcolor{crossset}0.09 &  & \cellcolor{crossset}87.7 & \cellcolor{crossset}90.3 & \cellcolor{crossset}0.84 & \cellcolor{crossset}0.13 \\
         &  & MDB-ss &  &
        \cellcolor{crossset}94.6 & \cellcolor{crossset}96.1 & \cellcolor{crossset}0.57 & \cellcolor{crossset}0.14 &  & \cellcolor{sameset}97.0 & \cellcolor{sameset}97.1 & \cellcolor{sameset}0.18 & \cellcolor{sameset}0.08 &  & \cellcolor{crossset}88.3 & \cellcolor{crossset}89.9 & \cellcolor{crossset}0.62 & \cellcolor{crossset}0.13 \\
         &  & PTDB &  &
        \cellcolor{crossset}95.6 & \cellcolor{crossset}96.9 & \cellcolor{crossset}0.63 & \cellcolor{crossset}0.13 &  & \cellcolor{crossset}96.3 & \cellcolor{crossset}96.6 & \cellcolor{crossset}0.21 & \cellcolor{crossset}0.09 &  & \cellcolor{sameset}89.7 & \cellcolor{sameset}91.2 & \cellcolor{sameset}0.56 & \cellcolor{sameset}0.13 \\
         &  & all &  &
        \cellcolor{multidataset}95.6 & \cellcolor{multidataset}96.7 & \cellcolor{multidataset}0.49 & \cellcolor{multidataset}0.14 &  & \cellcolor{multidataset}97.1 & \cellcolor{multidataset}97.3 & \cellcolor{multidataset}0.17 & \cellcolor{multidataset}0.08 &  & \cellcolor{multidataset}88.5 & \cellcolor{multidataset}90.0 & \cellcolor{multidataset}0.60 & \cellcolor{multidataset}0.13 \\
        \bottomrule
    \end{tabular}
    \begin{flushleft}
        \vspace{-2mm}
        \footnotemark[2] Results taken from the original paper~\citep{PENN}
    \end{flushleft}
\end{table*}

%% file: tables/speed_benchmark.tex
\begin{table}%[h]
\centering
\caption{Real-time factors (lower is better) for pitch detection models (RTF < 1 indicates real-time capable).}
\label{tab:rtf_comparison}
\normalsize{
\begin{tabular}{lcccc}
\toprule
 &  & \multicolumn{3}{c}{Real-time Factor} \\
\cmidrule{3-5}
Model &  & CPU &  & GPU \\
\midrule
PESTO &  & \textbf{0.0354} &  & \textbf{0.0032} \\
%PESTO-Stream &  & 0.0355 & 0.0032 \\
%\midrule
PENN &  & 0.1706 &  & 0.0096 \\
YIN &  & 0.0568 &  & - \\
\bottomrule
\end{tabular}}
\end{table}

%% file: tables/mirror.tex
\begin{table}%[ht!]
    \centering
    %\vspace{1.2cm}
    \caption{Effect of buffer refilling on PESTO's performance on the MIR-1k dataset. The model is trained and tested with VQT inputs modified according to the procedure detailed in Section~\ref{sec:lag}.
    A refill factor $m=0.0$ is equivalent to the regular PESTO model, while $m=0.5$ indicates the maximum possible refilling.}
    \label{tab:mirror}
    \resizebox{\columnwidth}{!}{
    % \small{
    \begin{tabular}{cccccccc}
        \toprule
         & & \multicolumn{6}{c}{Refill factor ($m$)} \\\cmidrule{3-8}
        Method & Metric & 0.0 & 0.1 & 0.2 & 0.3 & 0.4 & 0.5 \\
        \midrule
        \multirow{2}{*}{Refill}
        & RPA & 97.7 & 97.4 & 97.5 & 97.5 & 97.4 & 97.4 \\
        & RCA & 98.0 & 97.8 & 97.9 & 97.9 & 97.7 & 97.7 \\\midrule
        \multirow{2}{*}{Zero}
        & RPA & 97.7 & 97.0 & 97.5 & 97.5 & 97.4 & 95.4 \\
        & RCA & 98.0 & 97.4 & 97.9 & 97.9 & 97.8 & 95.7 \\
        \bottomrule
    \end{tabular}
    \label{tab:ablation}
    }
    % }
    %\vspace{0.6cm}
\end{table}

%% file: tables/background-rpa.tex
\begin{table}
    \centering
    \renewcommand{\thefootnote}{\fnsymbol{footnote}}
    \caption{Robustness of PESTO and other baselines to background music on the MIR-1K dataset, with various Signal-to-Noise ratios.}
    \label{tab:background}
    \small{
    \begin{tabular}{lcccccccc}
        \toprule
         &  & \multicolumn{7}{c}{Raw Pitch Accuracy} \\
        \cmidrule{3-9}
        Model &  & clean &  & $\SI{20}{dB}$ &  & $\SI{10}{dB}$ &  & $\SI{0}{dB}$ \\
        \midrule
        PESTO \\
        $\beta = 0$ &  &
        97.2 &  & 93.7 &  & 81.6 &  & 46.8 \\
        $\beta = \frac{1}{2}$ &  &
        98.1 &  & 97.9 &  & 95.8 &  & 79.7 \\
        $\beta = 1$ &  &
        97.1 &  & 96.7 &  & 94.0 &  & 78.9 \\
        \\
        $\beta \sim \mathcal{N}(0, \frac{1}{2})$ &  &
        98.0 &  & 97.6 &  & 94.9 &  & 79.3 \\
        $\beta \sim \mathcal{N}(0, 1)$ &  &
        98.1 &  & 97.8 &  & 95.6 &  & 82.5 \\
        \\
        $\beta \sim \mathcal{U}(0, \frac{1}{2})$ &  &
        \textbf{98.3} &  & \textbf{98.0} &  & \textbf{95.9} &  & 79.2 \\
        $\beta \sim \mathcal{U}(0, 1)$ &  &
        98.0 &  & 97.6 &  & 95.2 &  & 80.8 \\
        \midrule
        SPICE\footnotemark[2] &  &
        91.4 &  & 91.2 &  & 90.0 &  & 81.6 \\
        CREPE   &  & 97.5 &  & 97.1 &  & 95.3 &  & \textbf{85.8} \\
        PENN    &  & 90.6 &  & 81.0 &  & 51.1 &  & 20.9 \\
        \bottomrule
    \end{tabular}}
    \begin{flushleft}
        \vspace{-2mm}
        \footnotesize{\footnotemark[2] Results taken from the original paper~\citep{SPICE}}
    \end{flushleft}
    %\hspace{-5cm}
\end{table}

%% file: tables/ablation.tex
\begin{table*}[ht!]
    %\centering
    %\vspace{1.2cm}
    \caption{Respective contribution of various design choices of our model (losses, data augmentations, Toeplitz layer) on its performances.}
    \footnotesize{
    \begin{subtable}{0.55\textwidth}
    \caption{Training on MIR-1k}
    \begin{tabular}{cccccccccc}
        \toprule
         &  &  & \multicolumn{2}{c}{MIR-1k} & \multicolumn{2}{c}{MDB-ss} & \multicolumn{2}{c}{PTDB} \\
        $\LL{equiv}$ & $\LL{inv}$ & $\LL{SCE}$ & RPA & RCA & RPA & RCA & RPA & RCA \\
        \midrule
        %\xmark & \xmark & \xmark &  &  0.7 & 12.1 &  1.0 &  7.8 &  0.0 &  5.4 \\
        \xmark & \xmark & \cmark &  1.6 &  6.1 &  0.8 & 11.5 &  0.0 & 11.5 \\
        \xmark & \cmark & \xmark & 88.2 & 88.5 & 78.7 & 78.9 & 76.0 & 77.4 \\
        \cmark & \xmark & \xmark & 97.3 & 97.6 & 87.2 & 92.2 & 82.2 & 84.1 \\
        \\
        \xmark & \cmark & \cmark & 97.2 & 97.8 & \textbf{95.8} & \textbf{96.5} & \textbf{88.3} & 90.0 \\
        \cmark & \xmark & \cmark & 97.2 & 97.5 & 90.4 & 91.6 & 82.4 & 84.3 \\
        \cmark & \cmark & \xmark & 97.1 & 97.5 & 86.9 & 93.6 & 85.9 & 87.7 \\
        \\
        \cmark & \cmark & \cmark & \textbf{97.7} & \textbf{98.0} & 94.8 & 95.9 & 87.7 & \textbf{90.3} \\
        \midrule
        \multicolumn{3}{l}{no data augmentations} &
        97.2 & 97.6 & 93.8 & 94.6 & 88.0 & 90.1 \\
        \multicolumn{3}{l}{no Toeplitz layer} &
         1.3 &  7.1 &  1.5 &  5.7 &  0.0 & 12.5 \\
        \bottomrule
    \end{tabular}
    \label{tab:ablation_mir1k}
    \end{subtable}
    \hfill
    \begin{subtable}{0.415\textwidth}
    \caption{Training on MDB-stem-synth}
    \label{tab:ablation_mdb}
    \begin{tabular}{cccccccc}
        \toprule
         & \multicolumn{2}{c}{MIR-1k} & \multicolumn{2}{c}{MDB-ss} & \multicolumn{2}{c}{PTDB} \\
        $\LL{equiv}$ & RPA & RCA & RPA & RCA & RPA & RCA \\
        \midrule
        %\multicolumn{3}{l}{MDB-stem-synth} \\
        \xmark &
        \textbf{95.4} & \textbf{96.6} & \textbf{97.0} & \textbf{97.1} & \textbf{88.6} & \textbf{90.3} \\
        \cmark &
        94.6 & 96.1 & \textbf{97.0} & \textbf{97.1} & 88.3 & 89.9 \\
        \bottomrule
    \end{tabular}
    
    \vspace{1.05cm}

    \caption{Training on PTDB}
    \label{tab:ablation_ptdb}
    \begin{tabular}{cccccccc}
        \toprule
         & \multicolumn{2}{c}{MIR-1k} & \multicolumn{2}{c}{MDB-ss} & \multicolumn{2}{c}{PTDB} \\
        $\LL{equiv}$ & RPA & RCA & RPA & RCA & RPA & RCA \\
        \midrule
        \xmark &
        95.1 & \textbf{96.9} & 95.6 & 96.0 & \textbf{89.8} & \textbf{91.3} \\
        \cmark &
        \textbf{95.6} & \textbf{96.9} & \textbf{96.3} & \textbf{96.6} & 89.7 & 91.2 \\
        \bottomrule
    \end{tabular}
    \end{subtable}
    }
    %\label{tab:ablation}
\end{table*}